\documentclass[iop,revtex4]{emulateapj}
\usepackage{color}
\usepackage{times}
\usepackage{graphicx}
\usepackage{subfigure}
\usepackage{float}
\usepackage{booktabs}
\usepackage[breaklinks,colorlinks,citecolor=cyan,linkcolor=cyan]{hyperref}
\usepackage{txfonts}
\usepackage{epsfig}
\usepackage{lscape}
\usepackage{epstopdf}
\usepackage{CJK}
\usepackage{natbib} 
\def\tco{$^{13}$CO}
\def\ceo{C$^{18}$O}
\def\deg{^{\circ}}
\renewcommand{\arraystretch}{2.0}

\newcommand{\hii}{H~\textsc{ii}}
\newcommand{\msun}{$M_\odot$}
\newcommand{\lsun}{$L_\odot$}
\newcommand{\kms}{km~s$^{-1}$}

\newcommand{\cm}{cm$^{-2}$}
\journalinfo{HMSCs; Draft version \today }
\submitted{Draft Version, \today}
\shorttitle{High-Mass Starless Clumps: the Sample}
\shortauthors{Yuan et al.}
\begin{document}
\begin{CJK}{UTF8}{gbsn}
\nocite{*}

\title{High-mass Starless Clumps in the inner Galactic Plane: the Sample and Dust Properties}
\author{Jinghua Yuan (袁敬华)\altaffilmark{1}\footnotemark[\dag], 
	Yuefang Wu (吴月芳)\altaffilmark{2}, Simon P. Ellingsen\altaffilmark{3},
	Neal J. Evans II\altaffilmark{4,5},
	Christian Henkel\altaffilmark{6,7},\\
	 Ke Wang (王科)\altaffilmark{8},
	Hong-Li Liu (刘洪礼)\altaffilmark{1}, Tie Liu (刘铁)\altaffilmark{5}, 
	Jin-Zeng Li (李金增)\altaffilmark{1}, Annie Zavagno\altaffilmark{9}}
\affil{$^1$National Astronomical Observatories, Chinese Academy of Sciences, 
	20A Datun Road, Chaoyang District, \\ 
	Beijing 100012, China; jhyuan@nao.cas.cn;}
\affil{$^2$Department of Astronomy, Peking University, 100871 Beijing, China;}
\affil{$^3$School of Physical Sciences, University of Tasmania, Hobart, Tasmania, Australia;}
\affil{$^4$Department of Astronomy, The University of Texas at Austin, 2515 Speedway, 
	Stop C1400, Austin, TX 78712-1205, USA;}
\affil{$^5$Korea Astronomy and Space Science Institute 776, Daedeokdae-ro, 
	Yuseong-gu, Daejeon 34055, Korea;}
\affil{$^6$Max-Planck-Institut für Radioastronomie, Auf dem H\"{u}gel 69, 53121 Bonn, Germany}
\affil{$^7$Astron. Dept., King Abdulaziz University, P.O. Box 80203,
	Jeddah 21589, Saudi Arabia}
\affil{$^8$European Southern Observatory, Karl-Schwarzschild-Str. 2, 
	D-85748 Garching bei M\"{u}nchen, Germany}
\affil{$^9$Aix Marseille Universit, CNRS, LAM (Laboratoire d'Astrophysique de Marseille) UMR 7326, 
	F-13388, Marseille, France}
%\footnotetext[\dag]{jhyuan@nao.cas.cn}
\footnotetext[]{FITS images for the far-IR to sub-mm data, H$_2$ 
	column density and dust temperature maps of all the HMSC candidates
	are available at \url{https://yuanjinghua.github.io/hmscs.html}. 
	Codes used for this work are publicly available from 
	\url{https://github.com/yuanjinghua/HMSCs_cat}}
\begin{abstract}
	
	We report a sample of 463 high-mass starless clump (HMSC) 
	candidates within $-60\deg<l<60\deg$ and $-1\deg<b<1\deg$. 
	This sample has been singled out from 10861 ATLASGAL clumps.
	All of these sources are not associated with any known 
	star-forming activities collected in SIMBAD and 
	young stellar objects identified using color-based criteria.
	We also make sure 
    that the HMSC candidates have neither point sources at 24 
    and 70 \micron~nor strong 
    extended emission at 24 \micron. 
    Most of the identified HMSCs are infrared ($\le24$ \micron)
    dark and some are even dark at 70 \micron. Their 
    distribution shows crowding in Galactic spiral arms and toward
    the Galactic center and some well-known star-forming complexes.
    Many HMSCs are associated with large-scale filaments.
    Some basic parameters were attained from
    column density and dust temperature maps constructed 
    via fitting far-infrared and submillimeter continuum data to
    modified blackbodies. The HMSC candidates have 
    sizes, masses, and densities similar to clumps associated with 
    Class II methanol masers and \hii~regions, suggesting they 
    will evolve into star-forming clumps. More than
    90\% of the HMSC candidates have densities above some
    proposed thresholds for forming high-mass stars. 
    With dust temperatures and luminosity-to-mass ratios 
    significantly
    lower than that for star-forming sources, the HMSC candidates 
    are externally heated and genuinely at very early stages 
    of high-mass star formation.
    Twenty sources with equivalent radius $r_\mathrm{eq}<0.15$ pc
    and mass surface density $\Sigma>0.08$ g cm$^{-2}$ could be possible
    high-mass starless cores. Further investigations toward
    these HMSCs would undoubtedly shed light on 
    comprehensively understanding 
    the birth of high-mass stars.

\end{abstract}
\keywords{infrared: ISM -- ISM: clouds -- stars: formation -- stars: massive -- submillimeter: ISM}

\section{Introduction}

    High-mass stars, through mechanical and radiative 
    input, play crucial roles in the structural formation and evolution in galaxies. 
    They are also the primary contributor of chemical enrichment in space. However, the forming
    process of high-mass stars still remains a mystery \citep{2014prpl.conf..149T}. 
    
    The turbulent core and competitive accretion models have been proposed 
    as alternative scenarios for the formation of massive stars 
    \citep{2003ApJ...585..850M,2001MNRAS.323..785B}. In the turbulent core
    model, the final stellar mass is pre-assembled in the collapsing core, so this model
    requires the existence of high-mass starless cores. 
    In contrast, the competitive accretion model 
    predicts that high-mass stars begin as clusters of small cores with 
    masses peaked around the Jeans mass of the clump and
    there is no connection between the mass of its birth core and the final stellar mass.
    Discriminating between these two models requires identification of the youngest high-mass
    star formation regions to enable investigation of the initial conditions.

    %In the past decade, infrared dark clouds 
	%(IRDCs) have been extensively studied to characterize the initial physical 
	%and chemical properties of the earliest stages of high-mass star formation. 
	%Although \citet{2009A&A...505..405P} reported that more than 30\% IRDCs have 
	%no 24 \micron~counterparts, in-depth follow-up studies of IRDCs often reveal 
	%signs of ongoing star formation \citep[e.g., ][]{2007ApJ...668..348B,
	%2011ApJ...741..120R,2011ApJ...735...64W,2012ApJ...745L..30W,
	%2014MNRAS.439.3275W}. The lack of sample of bona-fide starless sources
    %is still a major bottleneck in the study of high-mass star formation.

    Starless clumps are the objects that fragment into dense starless cores ($r<0.15$ pc) which
    subsequently contract to form individual or bound systems of protostars 
    \citep{2014prpl.conf..149T}. 
    Despite many searches, there are only few 
    candidate high-mass starless cores known to date 
    \citep[e.g.,][]{2011ApJ...735...64W,2014MNRAS.439.3275W,2013ApJ...779...96T,
    	2014ApJ...796L...2C,2017ApJ...834..193K}.
    Recently, one of the most promising candidates, G028.37+00.07 (C1), was removed from
    the list because ALMA and NOEMA observations reveal protostellar 
    outflows driven by the core \citep{2016ApJ...828..100F,2016ApJ...821L...3T}.
    Although a high-resolution, deep spectral imaging survey is the 
    ultimate way to verify the starless nature, such a survey must start from a 
    systematic sample.
%    Previous 
%     studies suggest that high-mass starless cores 
%    are rare and a key question is whether enough
%    can be found to explain the initial 
%    mass function (IMF). Although three plausible 
%    candidates with masses of 60 \msun, 30 \msun, and 70 \msun~have been detected in 
%    IRDC G028.37+00.07 (C1), EGO G11.92-0.61, and IRDC G028.37+00.07 (C9),
%    \citep{2013ApJ...779...96T,2014ApJ...796L...2C,2016arXiv160906008K},
%    further observations are needed to testify their reliability. 
%    Recently reported outflows in IRDC G028.37+00.07 (C1) 
%    reduce their number to merely two \citep{2016ApJ...821L...3T}. 
    More candidate cores embedded
    within high-mass starless clumps are essentially required so that some statistical 
    understanding can be achieved. How dense clumps fragment is another key 
    question that must be answered to understand
    high-mass star formation, especially in
    the competitive accretion scenario. Although fragments with 
    super-Jeans masses have been revealed in some infrared dark clouds (IRDCs) 
    with on-going star-forming activity
    \citep[e.g.,][]{2011ApJ...735...64W,2014MNRAS.439.3275W,
    	2013A&A...553A.115B,2015A&A...581A.119B,2015ApJ...804..141Z}, 
    whether the high-mass starless clumps can fragment
    into Jeans-mass cores is still a key open question. 
    Therefore, 
    identification and investigations of high-mass
    starless clumps are essential to understanding the formation of high-mass stars and
    clusters.

    Recent Galactic plane surveys of dust continuum emission
    have revealed numerous dense structures at a wide range of evolutionary stages, 
    including the starless phases. 
    The \textit{APEX} Telescope Large Area Survey of the Galaxy 
    \citep[ATLASGAL,][]{2009A&A...504..415S}, the Bolocam Galactic Plane Survey 
    \citep[BGPS,][]{2011ApJS..192....4A,2011ApJ...741..110D}, 
    and the \textit{Herschel} Infrared  Plane Survey
    \citep[Hi-GAL,][]{2010PASP..122..314M} have revealed 
    extended dust emission at far-Infrared (IR) to sub-millimeter (submm), 
    while surveys at near-to-mid IR, such as
    the Galactic  
    Legacy Infrared Mid-Plane Survey Extraordinaire survey
    \citep[GLIMPSE,][]{2003PASP..115..953B,2009PASP..121..213C} and the 
    Galactic Plane Survey using the MIPS (MIPGSGAL), reveal emissions
    from warm/hot dust and young stars. The combination of these 
    surveys  provides
    an unprecedented opportunity to identify a large sample of candidate starless clumps
    
%    Three major surveys, which cover wavelengths from far-infrared (IR)
%    to millimeter, are the \textit{APEX} Telescope Large Area Survey of the Galaxy 
%    \citep[ATLASGAL,][]{2009A&A...504..415S}, the Bolocam Galactic Plane Survey 
%    \citep[BGPS,][]{2011ApJS..192....4A,2011ApJ...741..110D}, and 
%    the \textit{Herschel} Infrared Galactic Plane Survey
%    \citep[Hi-GAL,][]{2010PASP..122..314M}. The data collected by these surveys, combined with
%    some archival near- to mid-IR data, provide us 
%    an unprecedented opportunity to identify a large sample of candidate starless clumps.

    Recently, \citet{2012A&A...540A.113T}, \citet{2015MNRAS.451.3089T}, and 
    \citet{2016ApJ...822...59S} have identified 210, 667, and 2223 starless
    clump candidates in the longitude ranges $10^\circ<l<20^\circ$,
    $15^\circ<l<55^\circ$, and $10^\circ<l<65^\circ$, respectively.
    And these endeavors have revealed some properties of 
    early stages of star formation. However, these samples use criteria
    which do not allow them to reliably discriminate between low-mass and
    high-mass clumps and in some cases may have misidentified star-forming
    objects ({see Section \ref{sec-comp}}). Also, the covered longitude ranges are limited.
    
    In this work, we identify a  
    sample of high-mass starless clumps with better coverage 
    of Galactic longitude ($-60\deg<l<60\deg$)
     based on multiwavelength 
    data from the GLIMPSE, MIPSGAL, Hi-GAL and ATLASGAL surveys. The data
    used in this work are described in Section \ref{sec:data}.
    The identification procedure is described in Section \ref{sec:identification}.
    In Section \ref{sec:distance}, we outline distance estimation and 
    spatial distributions. Based on continuum data from 160 \micron~to 
    870 \micron, some dust parameters are derived in Section \ref{sect:dust}.
    More in-depth discussions and a summary of the findings are given in Sections 
    \ref{sec:discuss} and \ref{sec:conclusions}.
    
\section{Data}\label{sec:data}

    This work is based on data from several Galactic
    plane surveys covering wavelengths from mid-IR 
    to sub-millimeter (submm).  The sample of dense clumps from the ATLASGAL survey\footnotemark[1]
    \citep{2009A&A...504..415S}, provides the basis for our investigation.  The ATLASGAL survey mapped 
    420 square degrees of the Galactic plane between 
    $-80\deg<l<+60\deg$, using 
    the LABOCA camera on the \textit{APEX} telescope at 870 \micron~
    with a 19.$\!\!$\arcsec2 angular resolution. The astrometry of
    the data set and the derived source positions have
    been assumed to be the same as the pointing accuracy of
    the telescope which is $\sim2-3''$ 
    \citep{2013A&A...549A..45C}. The absolute flux density uncertainty
    is estimated to be better than 15\% \citep{2009A&A...504..415S}.
    Structures larger than 2.$\!$\arcmin5 have been filtered out 
    during the reduction of the raw data because the 
    emission from the atmosphere mimics that from extended astronomical
    objects. The final maps, gridded into 
    $3\deg\times3\deg$ tiles with a pixel size of 6\arcsec, are available
    from the project site\footnotemark[2]. 
    The average noise in the maps was determined from the $|b|<1\deg$ portions of
    the maps to be about 70 mJy beam$^{-1}$. On the basis of these maps
    two source catalogs have been produced using the 
    Gaussclump \citep{2014A&A...565A..75C} and SExtractor 
    \citep{2013A&A...549A..45C,2014A&A...568A..41U} 
    algorithms. The Gaussclump source catalog
    with 10861 sources has been optimized for  
    small-scale embedded structures (i.e., 
    nearby cores and distant clumps), with background emission 
    from molecular clouds removed 
    \citep{2014A&A...565A..75C}. On the other hand, 
    the 10163 compact sources extracted using the SExtractor algorithm 
    are representatives of larger-scale clump or cloud structures.
    In this work, we have used the Gaussclump source catalog 
    of \citet{2014A&A...565A..75C} as the parent sample to identify starless clumps. 

    Point source catalogs from GLIMPSE and MIPSGAL surveys have 
    been used to identify possible young stellar object (YSO)
    candidates associated with ATLASGAL clumps. Using 
    the IRAC instrument on board the \emph{Spitzer Space Telescope}
    \citep{2004ApJS..154....1W}, the GLIMPSE project surveyed the inner $130\deg$
    of the Galactic Plane at 3.6, 4.5, 5.8, and 8.0 \micron~
    with $5\sigma$ sensitivities of 0.2, 0.2, 0.4, and 0.4 mJy, respectively. 
    In addition to the images, the GLIMPSE survey performed 
    point-source photometry which, in combination with the 
    2MASS point source catalog \citep{2006AJ....131.1163S}, 
    provides photometric data of 
    point sources in seven infrared bands. The MIPSGAL survey used the 
    Multiband Infrared Photometer on \emph{Spitzer} \citep{2009PASP..121...76C} to
    map an area comparable to GLIMPSE at longer infrared wavelengths. Version 3.0 of 
    the MIPSGAL data includes mosaics from only the 24 \micron~band,
    with a sky coverage of $|b|<1\deg$ for $-68\deg<l<69\deg$
    and $|b|<3\deg$ for $-8\deg<l<9\deg$. The angular resolution
    and $5\sigma$ sensitivity at 24 \micron~ are 6\arcsec~and 
    1.7 mJy, respectively. The images and point source catalogs
    of both GLIMPSE and MIPSGAL are available at the InfraRed
    Science Archive (IRSA)\footnotemark[3].

    Far-IR data from the Hi-GAL survey have been used to
    further constrain the starless clump candidates and investigate their
    dust properties. Hi-GAL  
    is a key project of the \emph{Herschel Space Observatory}
    which mapped the entire Galactic plane 
    with nominal $|b| \le 1\deg$ (following the Galactic warp). 
    The Hi-GAL data were taken in fast scan mode 
    (60 arcsec~s$^{-1}$) using the PACS (70 and 160 
    \micron) and SPIRE (250, 350, and 500 \micron) 
    instruments in the parallel mode. The maps have been reduced with
    the ROMAGAL pipeline \citep{2011MNRAS.416.2932T}, an enhanced
    version of the standard \emph{Herschel} pipeline specifically
    designed for Hi-GAL. The effective angular resolutions are 10$.\!\!\arcsec$2, 
    13$.\!\!\arcsec$5, 18$.\!\!\arcsec$1, 24$.\!\!\arcsec$9, and 36$.\!\!\arcsec$4 at
    70, 160, 250, 350, and 500 \micron, respectively.

	\footnotetext[1]{The ATLASGAL project is a collaboration 
	between the Max-Planck-Gesellschaft, the European Southern 
	Observatory (ESO) and the Universidad de Chile. It includes 
	projects E-181.C-0885, E-078.F-9040(A), M-079.C-9501(A), 
	M-081.C-9501(A) plus Chilean data.}

    \footnotetext[2]{\url{http://atlasgal.mpifr-bonn.mpg.de/}}

    \footnotetext[3]{\url{http://irsa.ipac.caltech.edu}}

\section{The High-Mass Starless Clumps Catalog} \label{sec:identification}

    \begin{figure}
    \centering
    \includegraphics[width=0.45\textwidth]{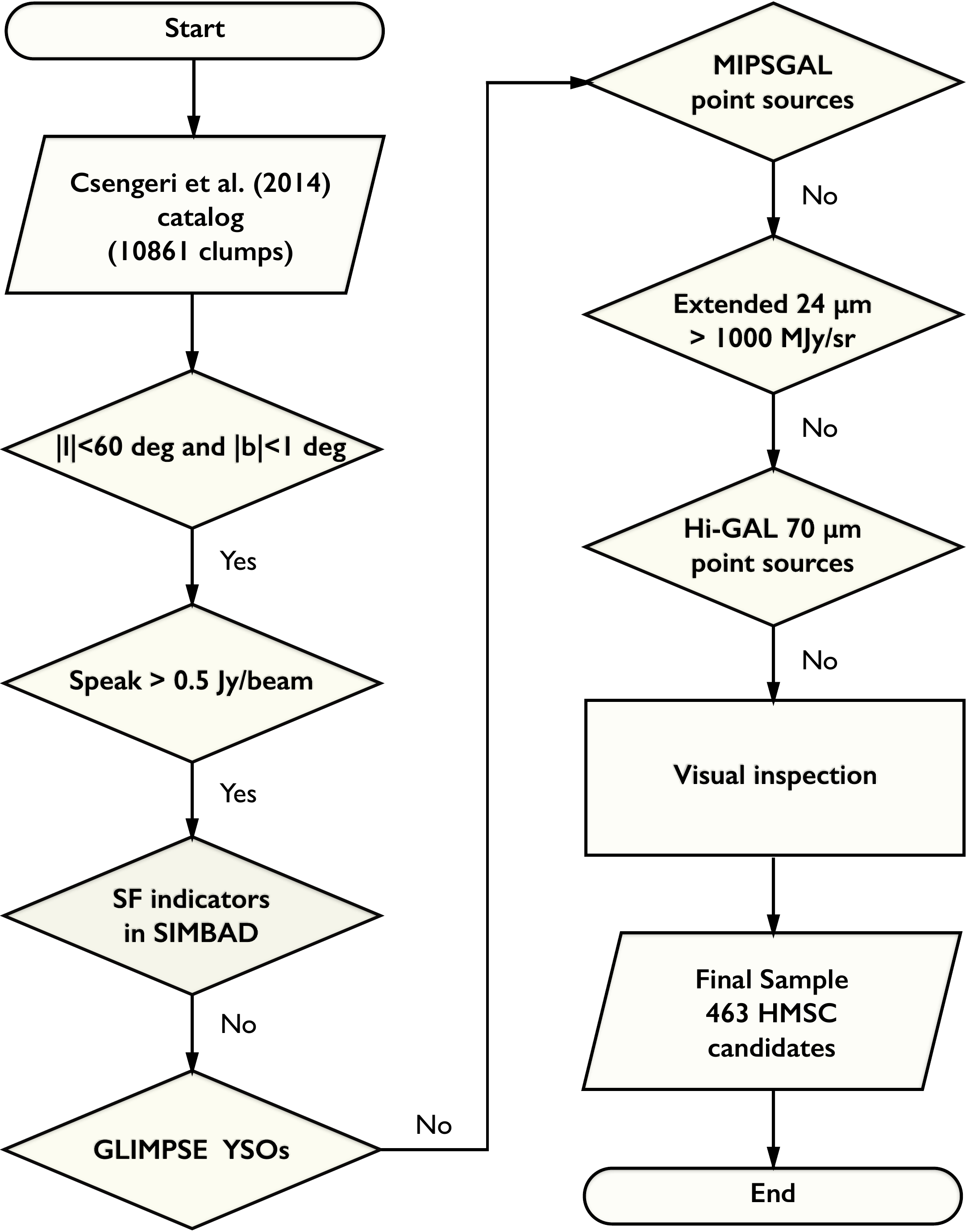}
    \caption{Flow chart describing the identification procedure 
		    of high-mass starless clumps from ATLASGAL compact 
		    sources.}\label{fig:FlowChart}
    \end{figure}

	\subsection{Source Identification}

    We have combined information from the GLIMPSE, MIPSGAL,
    Hi-GAL, and ATLASGAL surveys to identify 
    starless clumps. Reliable identification requires the combination of data
    from all four surveys and so only ATLASGAL 
    clumps in the inner Galactic plane ($|l|<60\deg$ and $|b|<1\deg$) which meet
    this criterion have been considered.  To identify candidate high-mass starless
    clumps we have applied the procedure outlined in \autoref{fig:FlowChart} to
    the ATLASGAL sources in the region of interest.

    %\subsection{Intensity Threshold at 870 \micron}

    As we mainly aim to find the birth-sites of 
    high-mass stars, a threshold for the peak intensity at 870
    \micron~is essential to identify high-mass
    clumps. Here, we followed \citet{2012A&A...540A.113T}
    and select only objects with peak intensities higher 
    than 0.5 Jy beam$^{-1}$. Assuming a dust opacity of $\kappa_{870}=1.65$ cm$^2$ g$^{-1}$
    and a dust temperature of 15 K, this threshold corresponds to 
    a column density of $2.2\times10^{22}$ \cm. 
    Although smaller than the theoretical threshold value proposed
    by \citet{2008Natur.451.1082K}, which is about 
    1 g \cm~or $2.15\times10^{23}$ \cm, we consider this to be a
    reasonable value considering the beam dilution which will occur
    due to the intermediate angular resolution 
    of the APEX observations. The effects of distance and
    telescope resolution on measurements of peak column density 
    have been studied by \citet{2009A&A...499..149V}
    under an assumed $r^{-1}$ density distribution.
    The peak column density at a spatial resolution of
    0.01 pc can be diluted by a factor of tens
    at $\sim$3 kpc if it is observed
    using a single dish telescope with an angular resolution of 
    $11$--$24$ arcsec \citep{2009A&A...499..149V}. Even assuming
    a crude correction factor of 10, clumps with
    peak intensities higher than 0.5 Jy beam$^{-1}$ likely 
    contain unresolved subregions with column densities higher than 
    $2.15\times10^{23}$ \cm. 

    \begin{deluxetable}{ll}
        \tablecaption{Star-forming indicators in 
        SIMBAD. \label{tb-simbad}}
        \tablewidth{0pt}
        
        \startdata
        \hline\hline
        Radio-source        &  Young Stellar Object \\
        centimetric Radio-source       & Young Stellar Object Candidate \\
        HII (ionized) region       &  Pre-main sequence Star \\
        Infra-Red source        & Pre-main sequence Star Candidate \\
        Far-IR source ($\lambda\ge30$ \micron)     & T Tau-type Star \\
        Near-IR source ($\lambda<10$ \micron)     & T Tau star Candidate \\
        Herbig-Haro Object      &  Herbig Ae/Be star \\
        Outflow     & Possible Herbig Ae/Be Star \\
        Outflow candidate     & Maser 
        \enddata
    \end{deluxetable}

    These first two criteria reduce the number of ATLASGAL sources under
    consideration by more than a factor of 2, with 5279 clumps remaining
    under consideration.  For these clumps we have used the SIMBAD database
    to search for a wide range of associated star-formation related phenomena.  We 
    queried the SIMBAD database for star-formation associated objects within a 
    circle defined by the major axis of the Gaussian ellipse 
    given in \citet{2014A&A...565A..75C}. If an object belonging to any of the 18 categories listed in 
    \autoref{tb-simbad} was found within this region, we consider the ATLASGAL source a 
    likely star-forming region and exclude it from our sample.
    This query to SIMBAD also successfully removes clumps associated with star-forming
    phenomena investigated in some large surveys such as the 6.7 GHz
    methanol masers \citep{2013MNRAS.435..524B}, the Red MSX Sources 
    \citep{2013ApJS..208...11L}, and \hii~regions \citep{2013MNRAS.435..400U}.
    
    We then further identify possible star-forming regions using data from the
    GLIMPSE point source catalog and applying the color criteria 
    given in \citet{2009ApJS..184...18G}. Any GLIMPSE source with colors 
    that fulfill the criteria below is considered as an YSO candidate and
    if any such sources lie within the region of the ATLASGAL clump it was excluded
    from our starless clump sample.
    $$
    [3.6]-[4.5]>0.7~\mathrm{and}~[4.5]-[5.8]>0.7~\mathrm{or~~~~~~~} 
    $$
    $$ \left\{
    \begin{array}{l}
     \left[3.6\right]-[4.5]-\sigma_1>0.15~\mathrm{and,}~ \\ 
     \left[3.6\right]-[5.8]-\sigma_2>0.35 ~\mathrm{and,}~ \\
     \left[4.5\right]-[8.0]-\sigma_3>0.5 ~\mathrm{and,}~ \\
     \left[3.6\right]-[5.8]+\sigma_2\le\frac{0.14}{0.04}
     \times(([4.5]-[8.0]-\sigma_3)-0.5)+0.5
    \end{array} \right. $$
    Here, $\sigma_1=\sigma([3.6]-[4.5])$, $\sigma_2=\sigma([3.6]-[5.8])$,  
    and $\sigma_3=\sigma([4.5]-[8.0])$ are combined errors, 
    added in quadrature.
    
    In a recent study, \citet{2013MNRAS.430..808G} show that 17\% of 
    the methanol masers from the 
    Methanol Multi-Beam (MMB) survey are not 
    associated with emission seen in GLIMPSE, indicating that 
    some very young high-mass stars may be too cold to be detectable in 
    IRAC bands but show weak 24 \micron~emission. 
    Thus, any clump which is 
    associated with a 24 \micron~point source probably hosts 
    star-forming activity and should be omitted. 
    By querying the 
    VizieR Service\footnotemark[4], we seized and 
    excluded clumps which are possibly associated with 
    24 \micron~point sources provided in \citet{2015AJ....149...64G}.
    Furthermore, some very faint point sources at 24 \micron~could 
    have been missed from the catalog of \citet{2015AJ....149...64G}
    due to bright extended emission contamination. In order to
    obtain a reliable sample of HMSCs, we also have ignored 
    clumps associated with 24 \micron~extended structures brighter
    than $10^3$ MJy sr$^{-1}$. {This threshold corresponds to
    about 2.0 mag within an aperture of $6\arcsec\!\!.35$, close
    to the 90\% differential completeness limit ($\sim1.98$ mag)
    of 24 \micron~point sources in bright and structured regions 
    \citep{2015AJ....149...64G}.}

	\footnotetext[4]{\url{http://vizier.u-strasbg.fr}}

    A total of 1215 ATLASGAL compact sources remain after applying
    these exclusion criteria.  These clumps were then subjected to a final visual inspection 
    to identify (and exclude) any which have very faint 24 \micron~point emission missed
    from the \citet{2015AJ....149...64G} catalog, any which are
    saturated at 24 \micron, and any which are associated with 70~\micron~
    point-like sources \citep{2016A&A...591A.149M}.

\begin{figure*}
    \centering
    \includegraphics[width=0.97\textwidth]{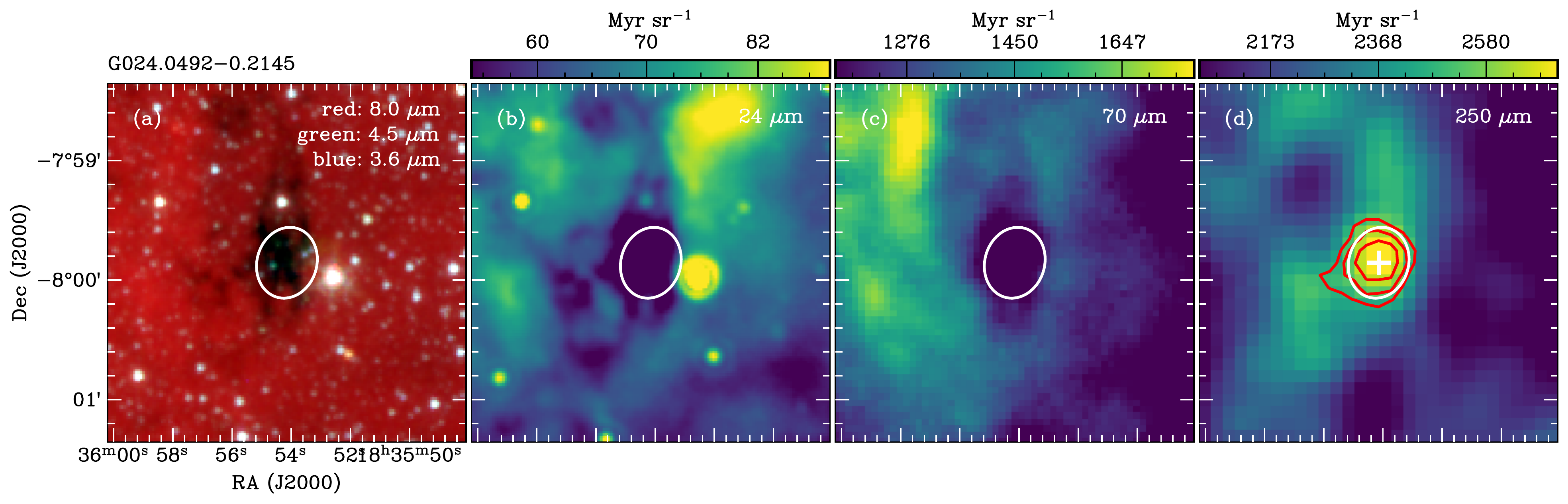}
    \includegraphics[width=0.97\textwidth]{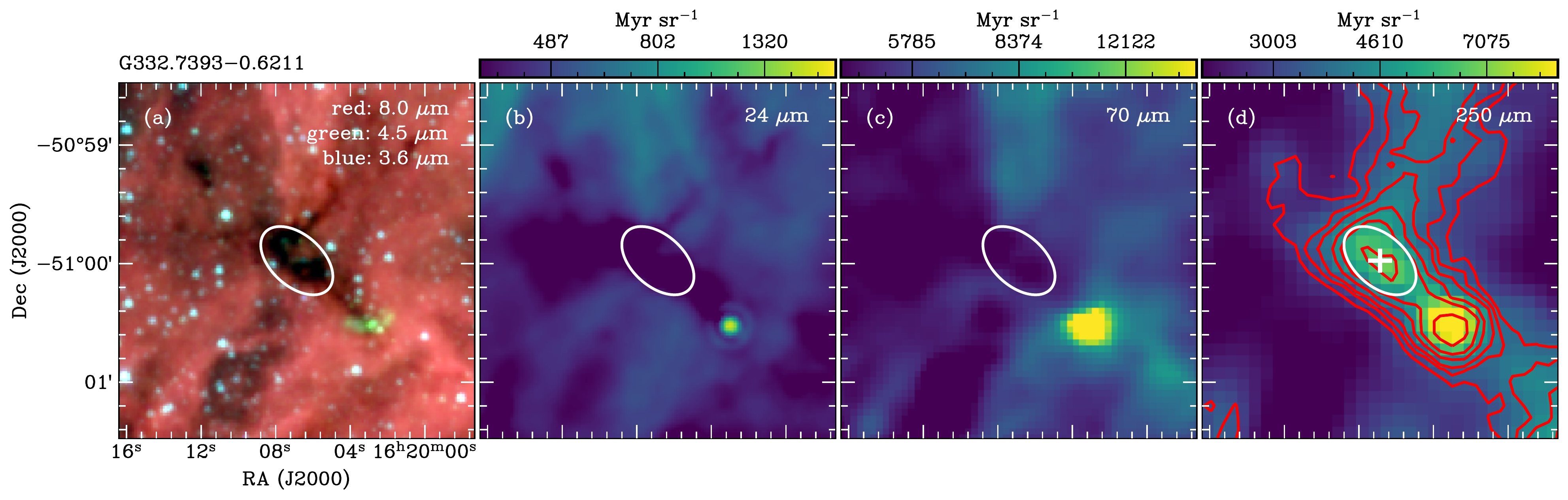}
    \caption{Morphology of two exemplary HMSCs in different wavelength bands. 
    	(a) Three color images with emission at 8.0, 4.5, and 3.6 
    	\micron~rendered in red, green, and blue, respectively.
    	(b),(c),(d) show dust continuum emission at 24, 
    	70, and 250 \micron, respectively. In panel (d),  
    	emission at 870 \micron~is presented in red 
    	contours with levels of 
    	0.3, 0.4, 0.5, 0.7, 0.9, 1.3, 1.8, 2.5, 4, 7
    	Jy beam$^{-1}$.
    	The white cross gives the peak position of 870 \micron, and 
    	the white ellipse delineates the source size 
    	based on the major and minor half-intensity 
    	axes provided in the ATLASGAL catalog.\\
    	(The complete figure set (463 images) is available.)
    }\label{fig:mulPlot}
    \end{figure*}

    {In the procedures of removing star-forming clumps, some star forming activities 
    could be in foreground, leading to  
    underestimating the number of starless clumps. However, the main goal of 
    this work is to identify a more reliable sample of HMSCs, not a complete catalog.}
    The resulting sample contains 463 HMSC candidates, which are listed 
    in \autoref{tb-cat}. We find that most of the HMSC candidates are associated
    with IRDCs, about 49\% (229/463) are even dark at 70 \micron~(see \autoref{fig:mulPlot}
    and column 11 of Table \ref{tb-cat}).
    Further inspection using far-IR continuum data supports the dense and cold nature of these sources 
    (see Sections \ref{sect:dust} and \ref{sec:discuss}).
    
    \subsection{Comparison with Reported Starless Clump Catalogs} \label{sec-comp}
    
    There are three previous papers that have searched
    for starless clumps. \citet{2012A&A...540A.113T} reported a sample of 
    210 starless clumps in the $10\deg<l<20\deg$ range. Star-forming clumps
    were ruled out via identifying YSOs from the GLIMPSE catalog 
    based on IR colors and visual analysis of 24 \micron~images 
    \citep{2012A&A...540A.113T}. Loose criteria for filtering out 
    star-forming clumps led to a greater than 
    50\% misidentification in the \citet{2012A&A...540A.113T} catalog
    \citep{2016ApJ...822...59S}.
    Among the 210 starless clumps of
    \citet{2012A&A...540A.113T}, only 107 have counterparts
    in the ATLASGAL GaussClump  Catalog \citep{2014A&A...565A..75C}. 
    And more than 70\% (75/107) are possible star-forming clumps if 
    diagnosed using the criteria applied in this work. 
    
    In the $15\deg<l<55\deg$ and $|b|<1\deg$ area, \citet{2015MNRAS.451.3089T}
    identified 667 starless clumps in IRDCs based on Hi-GAL data.
    The \texttt{Hyper} algorithm was used for clump extraction, and
    counterparts at 70 \micron~were used for protostellar identification.
    Spatial cross matching shows that 175 starless clumps from
    \citet{2015MNRAS.451.3089T} have counterparts in the ATLASGAL 
    catalog. And only 20 ($\sim11\%$) are associated with HMSC candidates identified
    in this work. The remaining 89\% (155/175)
    possibly can only form low-mass stars or already
    host star-forming activity. \citet{2015MNRAS.451.3089T} suggested that about 26\% 
    of their starless clumps have the potential to form high-mass stars. This
    suggests that about 50\% starless clumps reported in \citet{2015MNRAS.451.3089T} 
    may already have started forming stars.
    
    Another catalog of starless clumps has been compiled by 
    \citet{2016ApJ...822...59S} based on Bolocam survey data. Star-forming 
    indicators including mid-to far-IR YSOs, masers, and ultra-compact \hii~
    regions were considered to rule out clumps with current star formation. Among the
    2223 starless clumps with $10\deg<l<65\deg$ from \citet{2016ApJ...822...59S},
    179 have counterparts in the ATLASGAL catalog, and 35 ($20\%$) 
    are associated with HMSC candidates identified in this work. 
    As noted in \citet{2016ApJ...822...59S}, about 10\% 
    of their starless clumps have the potential to form high-mass 
    stars.
%    The fraction 
%    of sources identified by \citet{2016ApJ...822...59S} as starless clumps and 
%    our HMSC candidates is consistent with the fraction of 
%    high-mass stellar birth sites \citep{2016ApJ...822...59S}.
    
    In brief, we may have singled out the by far largest sample of 
    relatively reliable high-mass starless clump candidates distributed
    throughout the whole inner Galactic plane.

\section{Distances and Spatial Distribution} \label{sec:distance}

%    \begin{figure*}
%    \centering
%    \includegraphics[width=0.32\textwidth]{G012_hiSpec}
%    \includegraphics[width=0.32\textwidth]{G017_hiSpec}
%    \includegraphics[width=0.32\textwidth]{G033_hiSpec}
%    \caption{Example \hi~spectra. The black solid and red dashed lines in the upper
%    panel present the on- and off-source \hi~spectra (see the text for details). The 
%    difference between the on- and off-source \hi~spectra are presented in the 
%    lower panels. The systemic velocities are indicated with red vertical 
%    lines with cyan dash-dotted lines marking a 10 \kms~uncertainty. 
%    The blue dashed lines give the tangent velocities.
%    }\label{fig:hisa}
%    \end{figure*}

    \begin{figure*}
    \centering
    \includegraphics[width=0.95\textwidth]{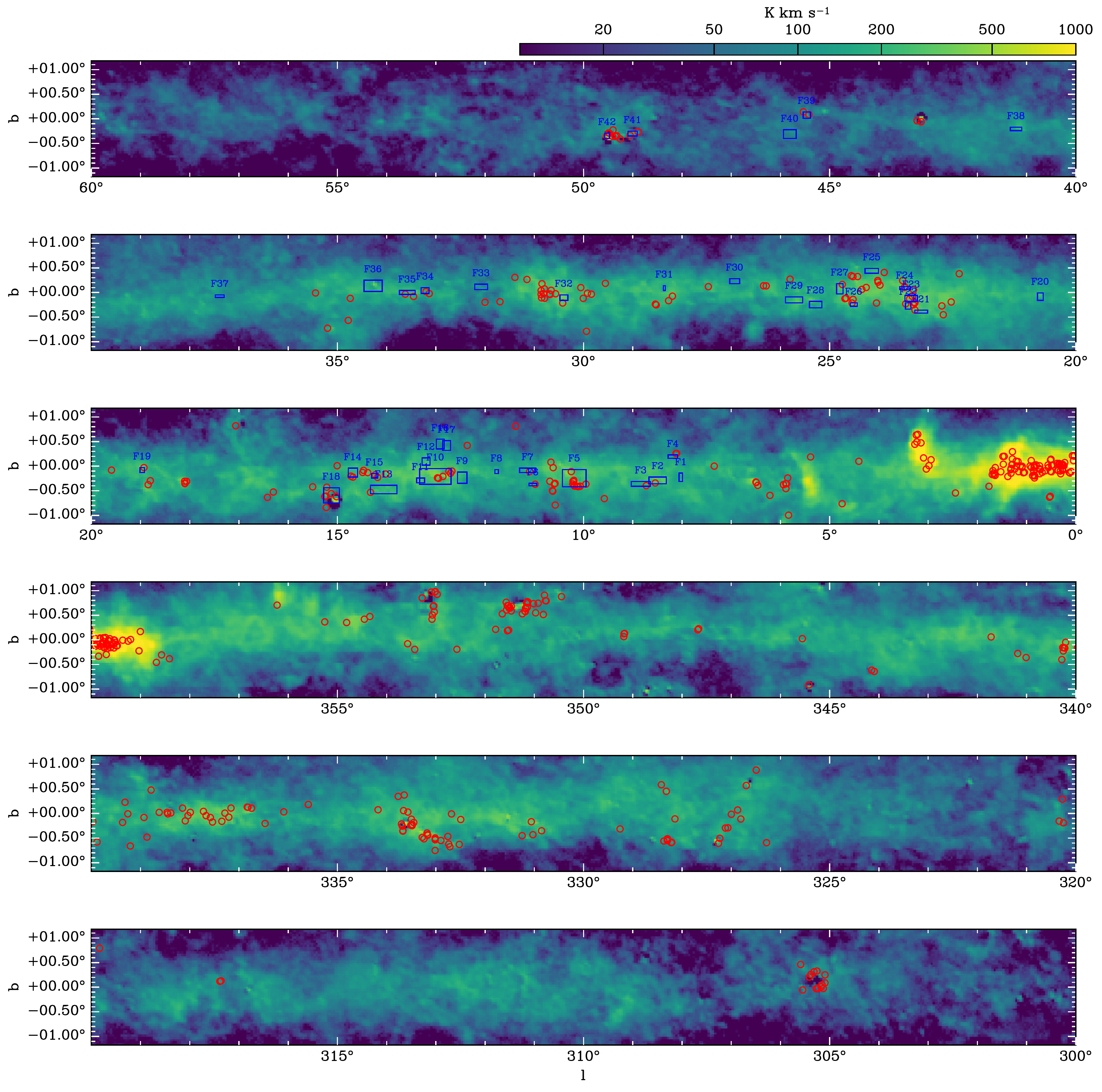}
    \caption{Distribution of HMSC candidates in the inner Galactic plane. 
    	The background shows CO $J=1-0$ emission from 
    	\citet{2014A&A...571A..13P}. The open boxes delineate 
    	the large scale dense filaments identified by 
    	\citet{2016ApJS..226....9W}.
    }\label{fig:distGP}
    \end{figure*}

    \begin{figure*}
    \centering
    \includegraphics[width=0.95\textwidth]{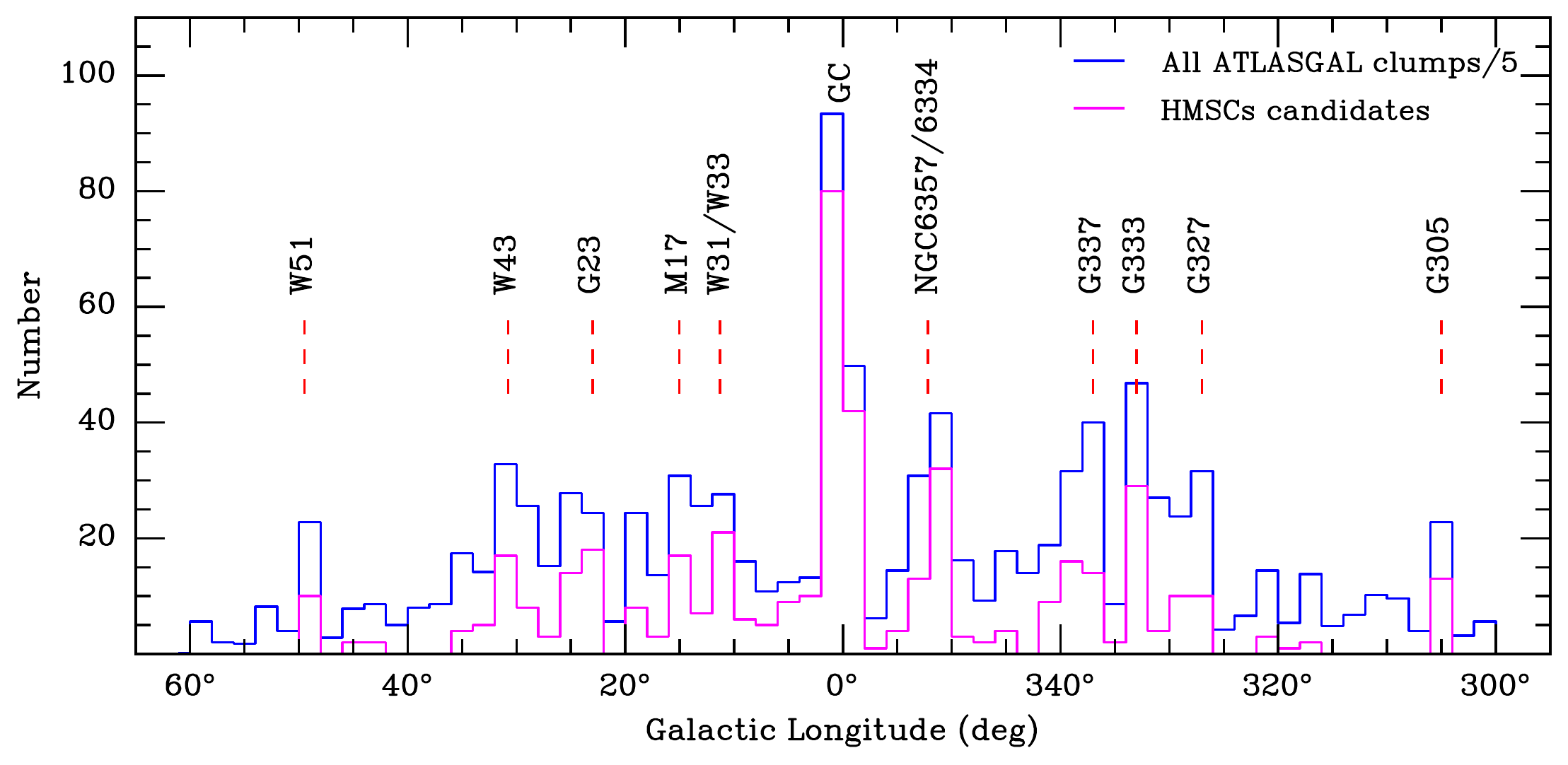}
    \caption{Histograms of Galactic longitudes of the HMSC candidates and all 
    	ATLASGAL clumps. Note that the number of sources in each bin 
    	for the full ATLASGAL catalog has been scaled down by 
    	a factor of five.}\label{fig:distL}
    \end{figure*}

    \begin{figure}
    \centering
    \includegraphics[width=0.46\textwidth]{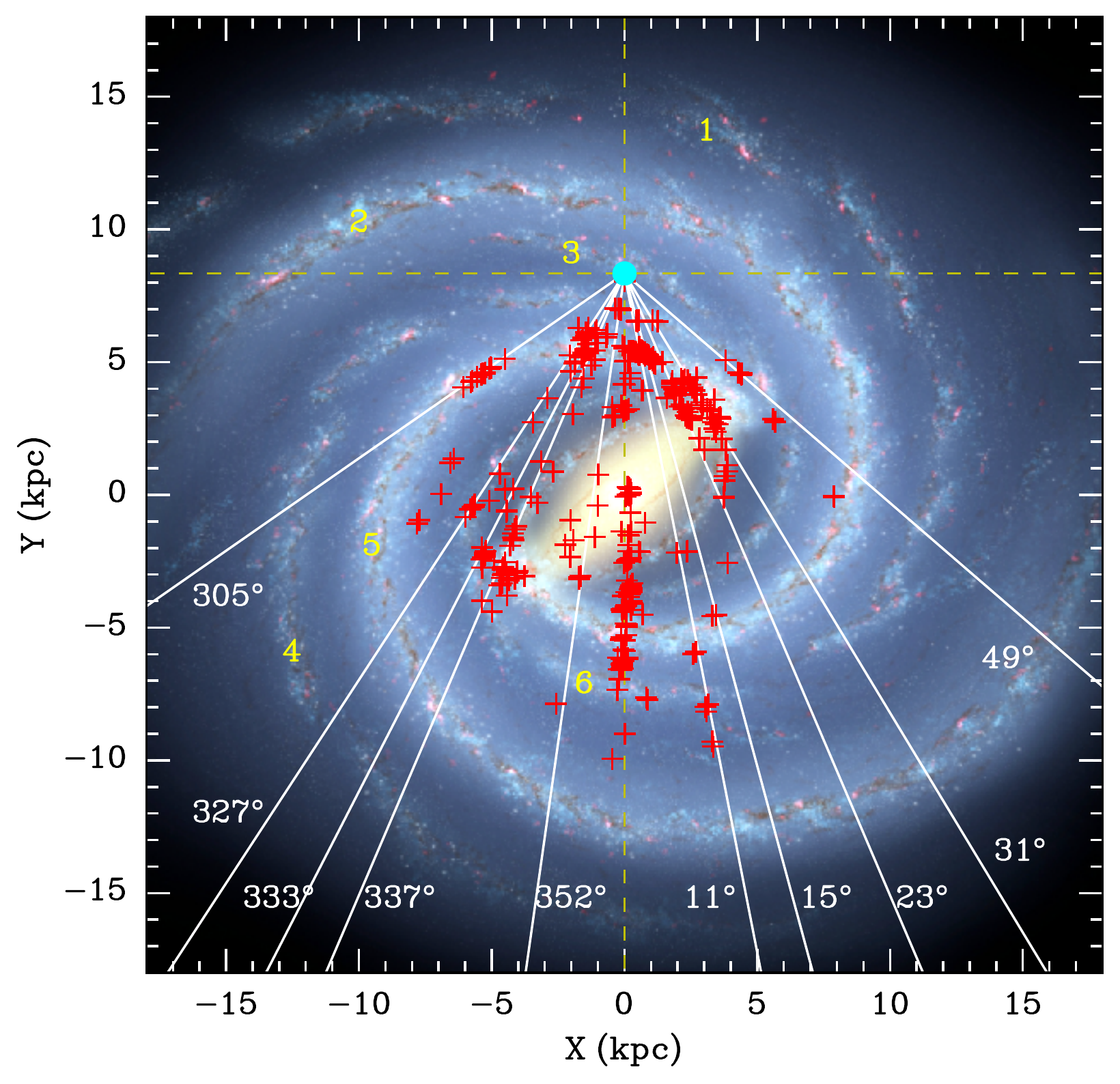}
    \caption{Spatial distribution of the identified HMSCs projected 
    	onto a top-down schematic of the Milky Way 
    	(artist's concept, R. Hurt: NASA/JPL-Caltech/SSC). 
    The spiral arms are indicated using numbers from 1 to 6, referring to
    the Outer, Perseus, Local, Carina-Sagittarius, Scutum-Centaurus, and Norma arms.\\
    (Codes and FITS files for making this plot are available from
	\url{https://github.com/yuanjinghua/Top_Down_Miky_Way}) }\label{fig:mw_faceOn}
    \end{figure}

	\begin{deluxetable*}{lcccccccccc}
		\tablecaption{Basic parameters of HMSC candidates. \label{tb-cat}}
		\tablewidth{0pt}
		\tablehead{
			\colhead{Designation} &\colhead{R.A.}& \colhead{Dec.}& \colhead{$\Theta_\mathrm{maj}$} &
			\colhead{$\Theta_\mathrm{min}$} & \colhead{PA\,\tablenotemark{a}} & \colhead{FWHM} & 
			\colhead{$V_\mathrm{lsr}$} & \colhead{Distance} &
			\colhead{Reference\,\tablenotemark{b}} &
			\colhead{70 \micron~dark\,\tablenotemark{c}}\\
			\colhead{} & \colhead{}  & \colhead{} & \colhead{(\arcsec)} & \colhead{(\arcsec)} &
			\colhead{($\deg$)} & \colhead{(\arcsec)} & 
			\colhead{(\kms)} &  \colhead{(kpc)}  
		}
		\startdata
		 G000.3404+0.0562 & 17h46m12.62s & -28d36m58.3s &    45 &    30 &      163 &    37 &   -11.80 &     4.19(0.22) &   JCO & Y \\
		G003.2278+0.4924 & 17h51m13.61s & -25d55m02.3s &    55 &    30 &       90 &    41 &    16.10 &     2.93(0.15) &   SMT & N \\
		G006.2130-0.5937 & 18h01m59.02s & -23d52m55.2s &    39 &    20 &      128 &    27 &    18.40 &     2.95(0.15) &   w12 & N \\
		G008.1102+0.2591 & 18h02m49.73s & -21d48m40.0s &    25 &    23 &        8 &    24 &    18.90 &     2.98(0.22) &   c14 & N \\
		G008.5441-0.3421 & 18h06m00.12s & -21d43m41.2s &    34 &    26 &      151 &    30 &    37.10 &     4.45(0.23) &   w12 & N \\
		G008.7264-0.3959 & 18h06m35.23s & -21d35m42.7s &    28 &    21 &      137 &    24 &    39.10 &     4.47(0.24) &   c14 & N \\
		G010.0676-0.4076 & 18h09m25.37s & -20d25m43.0s &    36 &    24 &      101 &    29 &    11.40 &     3.11(0.21) &   s13 & N \\
		G010.1065-0.4168 & 18h09m32.26s & -20d23m56.4s &    41 &    27 &       27 &    33 &    11.50 &     3.11(0.21) &   w12 & N \\
		G010.1839-0.4050 & 18h09m39.19s & -20d19m32.2s &    34 &    20 &       38 &    26 &    15.30 &     3.12(0.21) &   c14 & N \\
		G012.8572-0.2088 & 18h14m21.94s & -17d53m13.9s &    31 &    25 &      177 &    28 &    32.70 &     3.00(0.31) &   d13 & Y \\
		G018.8441-0.3758 & 18h26m41.30s & -12d41m11.8s &    60 &    34 &       14 &    45 &    61.00 &     3.60(0.26) &   s13 & Y \\
		G018.9295-0.0289 & 18h25m35.64s & -12d26m57.1s &    48 &    37 &       45 &    42 &    43.60 &     3.33(0.18) &   p12 & Y \\
		G320.2715+0.2920 & 15h07m56.06s & -57d54m33.5s &    33 &    21 &       44 &    26 &   -32.10 &    10.80(0.49) &   j08 & Y \\
		G320.3385-0.1534 & 15h10m04.01s & -58d15m38.5s &    24 &    20 &      121 &    22 &    -9.00 &    12.24(0.51) &   u14 & Y \\
		G326.4923+0.8820 & 15h42m37.99s & -53d58m02.6s &    32 &    24 &      108 &    28 &   -39.30 &     2.46(0.45) &   j08 & Y \\
		G333.0151-0.4964 & 16h20m48.00s & -50d43m01.2s &    47 &    26 &      101 &    35 &   -56.40 &     3.54(0.42) &   u07 & N \\
		G333.1639-0.4413 & 16h21m13.22s & -50d34m22.4s &    32 &    20 &       17 &    25 &   -52.60 &     3.36(0.43) &   c14 & N \\
		G336.4689-0.2023 & 16h34m13.27s & -48d01m30.0s &    47 &    22 &       62 &    32 &   -24.30 &    13.44(0.55) &   c14 & N \\
		G336.7428+0.1078 & 16h33m58.22s & -47d36m48.2s &    36 &    28 &       69 &    32 &   -76.30 &    10.66(0.41) &   m90 & N \\
		G359.9214+0.0276 & 17h45m19.51s & -28d59m20.0s &    40 &    40 &       59 &    40 &    59.30 &    10.90(0.24) & T13CO & N  
		\enddata
		\tablenotetext{a}{Position angle corrected in the FK5 system 
			with respect to the north direction.}
		\tablenotetext{b}{References for Systemic Velocity.~
			c14 = \citet{2014A&A...565A..75C};
			d11 = \citet{2011ApJ...741..110D};
			d13 = \citet{2013ApJS..209....8D};
			j08 = \citet{2008ApJ...680..349J};
			p12 = \citet{2012MNRAS.426.1972P};
			s13 = \citet{2013ApJS..209....2S};
			u07 = \citet{2007A&A...474..891U};
			u14 = \citet{2014MNRAS.437.1791U};
			w12 = \citet{2012A&A...544A.146W};
			m90 = MALT90;
			JCO = CO from the JCMT archive;
			T13CO = \tco~from the ThrUMMS survey;
			SMT = single-point observations using SMT.
		}
		\tablenotetext{c}{`Y' indicates the clump is associated with extinction feature at 70 \micron.}
		\tablenotetext{}{(This table is available in its entirety in machine-readable form.)}

	\end{deluxetable*}

    \subsection{Distance Estimation} \label{distance}

    The distance to a source is a fundamental parameter which is essential to determine
    its mass and luminosity. The distance to the HMSC candidates 
    is not known for most sources and to address this we have collected the systemic velocities
    from the literature for 294 clumps. 
    For further 20, 63, and 58 sources, velocity measurements
    	have been obtained based on data collected from the
    	MALT90 survey\citep{2013PASA...30...57J}, the ThrUMMS survey 
    	\citep{2015ApJ...812....6B} and
    	the JCMT archive. Velocities for the remaining 28
    	sources were determined via single-point observations of 
    	\tco~$(2-1)$ or \ceo~$(2-1)$ using the Submillimeter Telescope (SMT)
    	of the Arizona Radio Observatory (ARO). For all of the 
    	HMSCs, distance estimates were obtained using 
    a parallax-based distance estimator\footnotemark[5]. Leveraging results of 
	trigonometric parallaxes from the BeSSeL (Bar and Spiral Structure 
	Legacy Survey) and VERA (Japanese VLBI Exploration of Radio Astrometry) projects,
	the distance estimator can reasonably resolve kinematic distance
	ambiguities based on Bayesian approach \citep{2016ApJ...823...77R}.
	The distances calculated using the \texttt{FORTRAN} version of 
	the estimator are given in column 9 of Table \ref{tb-cat}.
	The 463 HMSCs have distances ranging from 0.3 to 18.3 kpc~with a mean of 7.1 kpc~and a 
	median of 6.0 kpc.

	\footnotetext[5]{\url{http://bessel.vlbi-astrometry.org/bayesian}}

    The equivalent radius ($r_\mathrm{eq}$) of the ATLASGAL clumps was estimated by multiplying  
    the distance by the deconvolved equivalent angular radius
    $\theta_\mathrm{eq} = \sqrt{\Theta_\mathrm{maj}\Theta_\mathrm{min}  
    	- \theta_\mathrm{HPBW}^2}$,
    where $\theta_\mathrm{HPBW}=19\arcsec\!\!.2$ is 
    the ATLASGAL beam. The major and minor half-intensity axes ($\Theta_\mathrm{maj}$ and $\Theta_\mathrm{maj}$)
    were obtained from \citet{2014A&A...565A..75C}. The resultant equivalent radii 
    are given in column 2 of Table \ref{tb-para}. The HMSC candidates  
    have a median equivalent radius of 0.65 pc, consistent with that of clumps identified 
    in some IRDCs \citep[$\sim0.6$ pc,][]{2015MNRAS.451.3089T} and 
    that of BGPS clumps \citep[$\sim0.64$ pc,][]{2016ApJ...822...59S}.

    \subsection{Spatial Distribution}\label{sec-spaceDis}

    The distribution of the identified HMSC candidates in the inner Galactic plane and as 
    a function of Galactic longitude are shown in 
    Figures \ref{fig:distGP} and  \ref{fig:distL}.
    The distribution of HMSCs in
    Galactic longitude is very similar to that of the 
    full sample of ATLASGAL clumps. A prominent overdensity can be observed 
    toward the Galactic center region. In addition, there are several
    well known star-forming complexes (i.e., W51, W43, G23, 
    M17, W31/W33, NGC 6334/6357, G337, G333, G327, and G305) standing out with significant peaks. 
	{Many of the HMSCs are associated with large-scale filaments identified by 
	\citet{2016ApJS..226....9W}. In the longitude range $7.5\deg<l<60\deg$ covered by both 
	\citet{2016ApJS..226....9W} and this study, there are 145 HMSCs, 
	39 of which are located on large-scale 
	filaments. This implies about 27\% (39/145) of future high-mass star formation
	would take place in large-scale ($>10$ pc) filamentary structures in the Galactic plane.}
	
    To determine the locations of these HMSCs in our Galaxy,  
    we have plotted
    the sources on a top-down schematic of the Milky Way (see
    Figure \ref{fig:mw_faceOn}). Most targets are located 
    in spiral arms in the inner Galaxy with Galactocentric
    distances $R_\mathrm{gal}<8.34$ kpc. 
    Most sources in the general direction of W51 ($l\sim49^\circ$) are located 
    in the Carina-Sagittarius arm, with an average distance of about 6 kpc
    consistent with that of the W51 complex \citep{2010ApJ...720.1055S}.
    Sources toward W43 ($l\sim31^\circ$) and G23 ($l\sim23^\circ$) mostly
    reside in the Scutum-Centaurus arm and some in the near 3 kpc arm. 
    The overdensities at $l=10-17^\circ$ are associated with 
    the W31 and W33 complexes and the well known star-forming region
    M17. The sources in this Galactic longitude range are mostly 
    located in the Scutum-Centaurus arm, with some in the Perseus and 
    Norma arms with distances larger than 10 kpc. The peak at 
    $l=352^\circ$ originates from two groups of sources with the near group 
    associated with the relatively nearby complex of NGC 6334/6357
    \citep{2010A&A...515A..55R} located in the Carina-Sagittarius
    arm, and the far group located in the Galactic Bar. The peaks toward
    G327, G333, and G337 are mostly located in the Scutum-Centaurus arm,
    and some in the Perseus, Norma and near 3 kpc arms. The sources in the Crux-Scutum
    arm at this direction have an average distance of about 3.5 kpc
    consistent with that of the G333.2-0.4 giant molecular cloud 
    \citep{2012MNRAS.419..211S}. Sources at $l=305\deg$ are mostly located
    at the tangent point of the Crux-Scutum arm.
%     with distances of about 4 kpc
%    consistent with the recently reported parallax distance 
%    to some of the methanol masers in the G305 complex \citep{2017MNRAS.465.1095K}.

    \begin{center}
    \begin{deluxetable*}{lcccccccccc}
    	\tablecaption{Physical parameters of HMSC candidates. \label{tb-para}}
    	\tablewidth{0pt}
    	\tablehead{
    		\colhead{Designation} & \colhead{$r_\mathrm{eq}$}& 
    		\colhead{$T_\mathrm{dust}$}& \colhead{$N_\mathrm{H_2}$}& 
    		\colhead{$n_\mathrm{H_2}$}& 
    		\colhead{$\Sigma_\mathrm{mass}$}&  \colhead{$M_\mathrm{cl}$}& 
    		\colhead{$L_\mathrm{cl}$}& \colhead{$L_\mathrm{cl}/M_\mathrm{cl}$}\\
    		\colhead{} &  \colhead{(pc)}  & \colhead{(K)}  & 
    		\colhead{($10^{22}$ cm$^{-2}$)}  & \colhead{($10^{4}$ cm$^{-3}$)}  & 
    		\colhead{(g cm$^{-2}$)}  & \colhead{($M_\odot$)}  & 
    		\colhead{($L_\odot$)} & \colhead{($L_\odot/M_\odot$)} 
    	}
    	\startdata
    	 G000.3404+0.0562 &     0.64 &    17.08 &    18.44 &     3.26 &     0.40 & 2.44e+03 & 4.83e+03 &     1.98 \\
    	G003.2278+0.4924 &     0.51 &    14.37 &     3.14 &     0.61 &     0.06 & 2.34e+02 & 1.88e+02 &     0.80 \\
    	G006.2130-0.5937 &     0.29 &    15.75 &     3.49 &     2.36 &     0.13 & 1.67e+02 & 2.44e+02 &     1.46 \\
    	G008.1102+0.2591 &     0.21 &    18.41 &     2.72 &     5.13 &     0.21 & 1.33e+02 & 4.76e+02 &     3.58 \\
    	G008.5441-0.3421 &     0.49 &    12.32 &     4.23 &     1.25 &     0.12 & 4.27e+02 & 1.41e+02 &     0.33 \\
    	G008.7264-0.3959 &     0.32 &    12.52 &     6.88 &     7.86 &     0.49 & 7.54e+02 & 2.48e+02 &     0.33 \\
    	G010.0676-0.4076 &     0.34 &    18.78 &     2.04 &     1.13 &     0.07 & 1.24e+02 & 5.48e+02 &     4.43 \\
    	G010.1065-0.4168 &     0.41 &    18.38 &     2.71 &     0.82 &     0.06 & 1.64e+02 & 6.05e+02 &     3.69 \\
    	G010.1839-0.4050 &     0.27 &    22.43 &     1.77 &     1.55 &     0.08 & 8.55e+01 & 1.20e+03 &    13.98 \\
    	G012.8572-0.2088 &     0.29 &    17.35 &     5.31 &     3.65 &     0.21 & 2.67e+02 & 6.60e+02 &     2.47 \\
    	G018.8441-0.3758 &     0.71 &    14.30 &     4.38 &     0.58 &     0.08 & 6.09e+02 & 5.12e+02 &     0.84 \\
    	G018.9295-0.0289 &     0.61 &    18.18 &     3.22 &     0.52 &     0.06 & 3.36e+02 & 1.40e+03 &     4.16 \\
    	G320.2715+0.2920 &     0.94 &    12.15 &     2.87 &     0.74 &     0.13 & 1.81e+03 & 5.19e+02 &     0.29 \\
    	G320.3385-0.1534 &     0.63 &    16.00 &     4.16 &     4.82 &     0.58 & 3.43e+03 & 5.52e+03 &     1.61 \\
    	G326.4923+0.8820 &     0.24 &    13.88 &     8.22 &     7.26 &     0.33 & 2.85e+02 & 1.76e+02 &     0.62 \\
    	G333.0151-0.4964 &     0.50 &    17.60 &     5.45 &     1.60 &     0.16 & 5.86e+02 & 1.37e+03 &     2.34 \\
    	G333.1639-0.4413 &     0.27 &    18.91 &     6.84 &     7.31 &     0.38 & 4.10e+02 & 1.69e+03 &     4.13 \\
    	G336.4689-0.2023 &     1.68 &    14.19 &     2.25 &     0.21 &     0.07 & 2.85e+03 & 2.20e+03 &     0.77 \\
    	G336.7428+0.1078 &     1.31 &    22.16 &     1.54 &     0.16 &     0.04 & 1.02e+03 & 1.42e+04 &    13.86 \\
    	G359.9214+0.0276 &     1.85 &    18.56 &     6.24 &     0.32 &     0.11 & 5.88e+03 & 2.09e+04 &     3.56   
    	\enddata
    	\tablenotetext{}{(This table is available in its entirety in machine-readable form.)}
    \end{deluxetable*}
	\end{center}
    %\end{landscape}
    
    \begin{figure}
    \centering
    \includegraphics[width=0.48\textwidth]{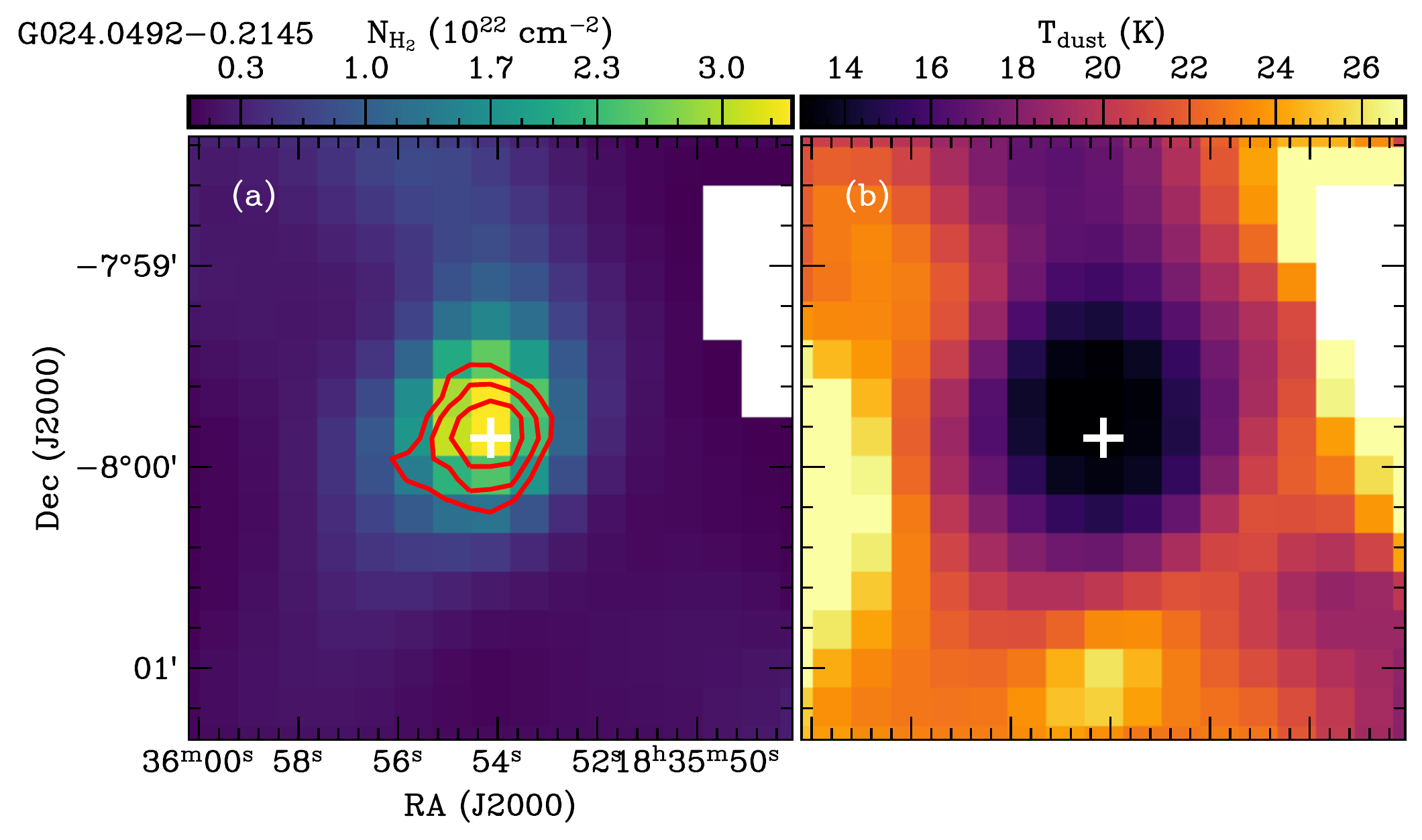}
    \includegraphics[width=0.48\textwidth]{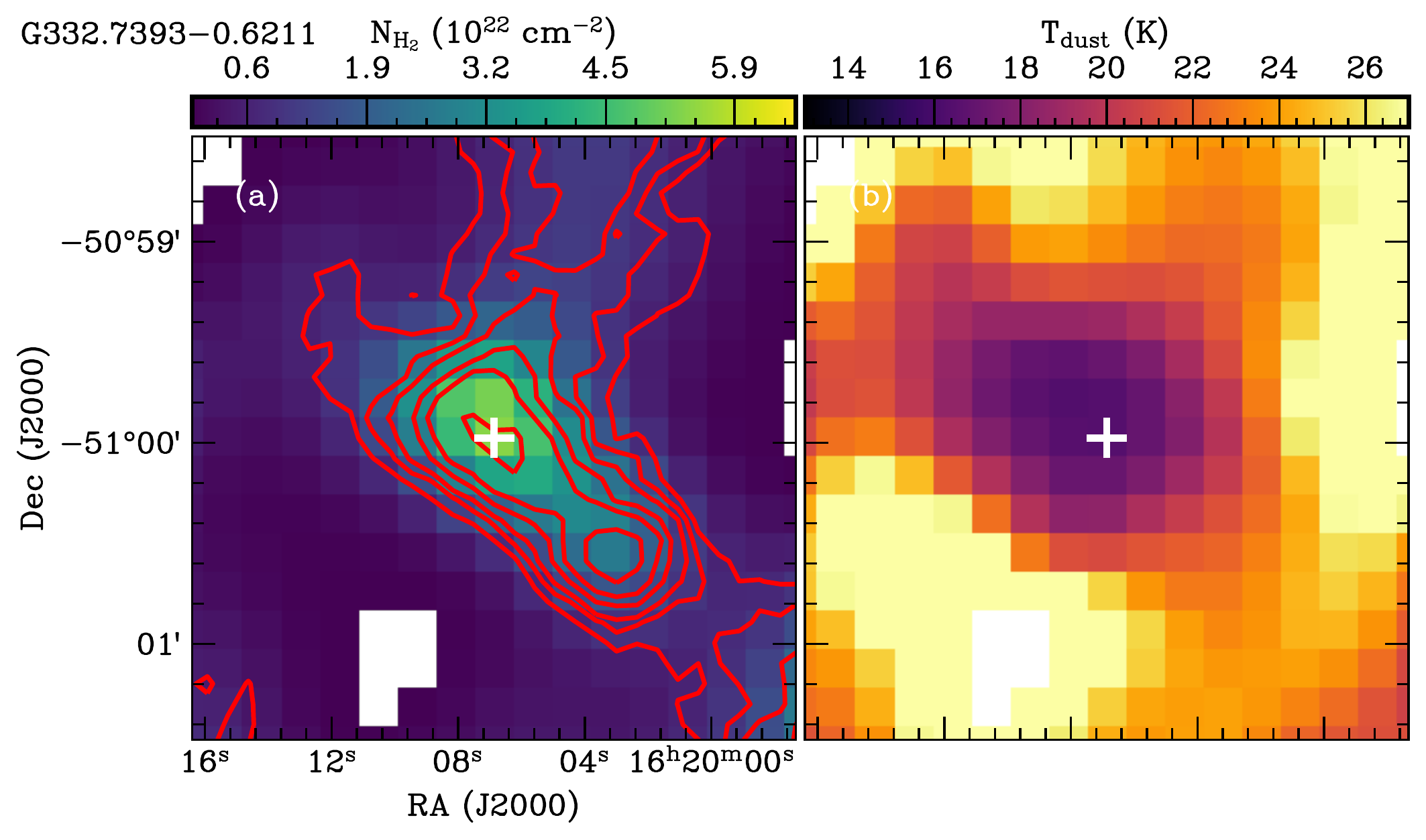}
    \caption{Column density (a) and dust temperature maps (b) 
    	of two exemplary HMSC candidates. The contours are the same to
	    the ones in Figure \ref{fig:mulPlot} (d).\\
	    (The complete figure set (463 images) is available.)
	    }\label{fig:sedImshow}
    \end{figure}

	\section{Dust Properties} \label{sect:dust}
	
    For all the 463 HMSC candidates, there are image data at six far-IR to 
    submm bands covering wavelengths from 70 to 870 \micron~
    collected by the Hi-GAL and ATLASGAL surveys, 
    enabling us to obtain some physical parameters. 
    As these sources effectively do not emit at 70 \micron, 
    we only take emission at 160, 250, 350, 500, and 870 \micron~into
    account to obtain physical parameters.

    \subsection{Convolution to a Common Resolution and Foreground/Background Filtering}

    In order to be able to estimate physical properties from multiwavelength
    observations with different angular resolutions, we first convolved the images
    to a common angular resolution of $36.\!\!\arcsec4$ which is essentially
    the poorest resolution of any of the wavelengths. 
    The \textit{convolution} package of 
    \textit{Astropy} \citep{2013A&A...558A..33A} was used with a Gaussian kernel
    of $\sqrt{36.\!\!\arcsec4-\theta_\lambda^2}$, where $\theta_\lambda$
    is the HPBW beam size for a given Hi-GAL or ATLASGAL band. Then the convolved data from the
    different bands were
    re-gridded to be aligned pixel-by-pixel with a common pixel size of $11.\!\!\arcsec5$.

    As our targets are in the Galactic plane, foreground/background removal 
    is essential to reduce the uncertainties in some of the derived parameters.
    For the ATLASGAL images, any uniform astronomical signal on spatial 
    scales larger than $2.\!\arcmin5$ has been filtered out 
    together with atmospheric emission during 
    the data reduction \citep{2009A&A...504..415S}. The filtering of 
    Hi-GAL images was performed using the \textit{CUPID-findback} algorithm
    of the \textit{Starlink} suite\footnotemark[6]. The algorithm constructs 
    the background iteratively from the original image. 
    At first, a filtered form of the input data is produced 
    by replacing every input pixel by the minimum of the 
    input values within a rectangular box centered on the pixel. 
    This filtered data is then filtered again, using a filter 
    that replaces every pixel value by the maximum value in a 
    box centered on the pixel. Then each pixel in this filtered 
    data is replaced by the mean value in a box centered on the pixel. 
    The same box size is used for the first three steps. 
    The final background estimate is obtained via some corrections 
    and iterations through comparison with the initial input data. 
    For further details on the algorithm please see the  
    online document for \textit{findback}\footnotemark[7]. 
    As a key parameter, the filtering box was chosen to be $2.\!\arcmin5$
    for consistency with the ATLASGAL data. One drawback
    of the \textit{findback} algorithm is that the background can be
    overestimated when processing high signal-to-noise data, especially
    when there are strong emission features significantly smaller than 
    the box size. To mitigate issues with overestimation of the background level, we 
    set the parameter NEWALG to TRUE and iteratively ran \textit{findback}.
    The background image resulting from the previous run was used
    as the input data for the next iteration. After careful examination of a number of test sources,
    we found that the resulting background images became stable after
    five iterations. The background images after five iterative runs of 
    \textit{findback} were subtracted from the post-convolution data 
    for all Hi-GAL bands to remove large-scale structures.

    \footnotetext[6]{\url{http://starlink.eao.hawaii.edu/starlink/WelcomePage}}
    \footnotetext[7]{\url{http://starlink.eao.hawaii.edu/starlink/findback.html}}
      
    \subsection{Spectral Energy Distribution Fitting}

    We have used the smoothed and background-removed far-IR to submm image data to 
    obtain intensity as a function of wavelength for each pixel and applied a modified blackbody
    model to this data.
    \begin{equation}\label{eq-gb}
    I_\nu=B_\nu(T) (1-e^{-\tau_\nu})
    \end{equation}
    where the Planck function $B_\nu(T)$ is modified by optical depth
    \begin{equation}
    \tau_\nu = \mu_\mathrm{H_2}m_\mathrm{H}\kappa_\nu N_\mathrm{H_2}/R_\mathrm{gd}.
    \end{equation}
    Here, $\mu_\mathrm{H_2}=2.8$ 
    is the mean molecular weight adopted from \citet{2008A&A...487..993K},
    $m_\mathrm{H}$ is the mass of a 
    hydrogen atom, $N_\mathrm{H_2}$ is the H$_2$ column density, $R_\mathrm{gd}=100$ is 
    the gas to dust ratio. The dust opacity $\kappa_\nu$  
    can be expressed as a power law in frequency,
    \begin{equation}
    \kappa_\nu=3.33\left(\frac{\nu}{600~\mathrm{GHz}}\right)^\beta~\mathrm{cm^2g^{-1}}.
    \end{equation} 
    where $\kappa_\nu(\mathrm{600~GHz})=3.33~\mathrm{cm^2g^{-1}}$ is the dust opacity for
    coagulated grains with thin ice mantles  
    \citep[from column 5 of table 1 in][ often referred to as OH5]{1994A&A...291..943O},
    but scaled down by a factor of 1.5 as suggested in 
    \citet{2010ApJ...712.1137K}. The scaled OH5 dust opacities are 
    consistent with the values used in other high-mass star-formation studies 
    \citep[e.g.,][]{2008A&A...487..993K,2010ApJ...723L...7K,
    	2010A&A...518L..97E,2013ApJ...772...45E,
    	2013A&A...549A.130V,2015MNRAS.451.3089T}.
    The dust emissivity index has been
    fixed to be $\beta=2.0$ in agreement with the standard value for cold
    dust emission \citep{1983QJRAS..24..267H}. The 
    free parameters in this model are the dust temperature and the column density. 

    The fitting was performed using the Levenberg-Marquardt algorithm provided in
    the python package \textit{lmfit}\footnotemark[8] \citep{2016ascl.soft06014N}.
    Only pixels with positive intensities in the four 
    longest wavelength Hi-GAL bands and the 
    ATLASGAL band were modeled and the inverse of the rms 
    errors in the images were used as weights in the fit.
    We found that pixels with 60 \micron~intensities $<60$ MJy sr$^{-1}$ 
    (about $3\sigma$) cannot be well fitted. For these
    pixels, only the data at wavelengths greater than or 
    equal to 250 \micron~were used.

	\footnotetext[8]{\url{https://lmfit.github.io/lmfit-py/index.html}}
	
    The resultant column density and dust temperature maps
    are presented in \autoref{fig:sedImshow}. Most HMSC candidates show coincidence 
    between density maximum and temperature minimum, as is seen in the sources presented in \autoref{fig:sedImshow}. This is 
    in line with the absence of detectable star-forming activity. 

    \subsection{Physical Parameters}

    The beam-averaged column densities and dust temperatures at the 
    peak positions were extracted from the relevant maps and are listed in \autoref{tb-para}.

    We integrated the column densities 
    in the scaled ellipses at a resolution of $36.\!\!\arcsec4$~to 
    estimate the clump mass using the relationship,
    \begin{equation}
        M_\mathrm{clump} = 
        \mu_\mathrm{H_2}m_\mathrm{H}d^2\Omega_\mathrm{pix}\sum{N_\mathrm{H_2}}.
    \end{equation}
    Here, $d$ is the source distance and $\Omega_\mathrm{pix}$ is the solid
    angle of one pixel.
    The major and 
    minor axes of the scaled ellipses were obtained via 
    $\Theta^\mathrm{maj}_\mathrm{36.4} = 
    \sqrt{{\Theta^\mathrm{maj}_\mathrm{atl}}^2-{19.\!\!\arcsec2}^2+{36.\!\!\arcsec4}^2}$ and
    $\Theta^\mathrm{min}_\mathrm{36.4} = 
    \sqrt{{\Theta^\mathrm{min}_\mathrm{atl}}^2-{19.\!\!\arcsec2}^2+{36.\!\!\arcsec4}^2}$, where
    $\Theta^\mathrm{maj}_\mathrm{atl}$ and $\Theta^\mathrm{min}_\mathrm{atl}$
    are the major and minor axes given in the ATLASGAL catalog. The source average
    $\mathrm{H_2}$ number densities were calculated to be
    \begin{equation}
    n_\mathrm{H_2} = \frac{M_\mathrm{clump}}{(4/3)\pi r_\mathrm{eq}^3
    \mu_\mathrm{H_2}m_\mathrm{H}}.
    \end{equation}
    The mass surface densities were derived via
    \begin{equation}
    \Sigma_\mathrm{mass} = \frac{M_\mathrm{clump}}{\pi r_\mathrm{eq}^2}.
    \end{equation}
    Here, $r_\mathrm{eq}$ is the equivalent physical radius.

    In the process of fitting the data, we determined the frequency-integrated 
    intensity ($I_\mathrm{int}$) for each pixel using the resultant dust temperature and 
    column density. The luminosities of the sources with distance measurements
    were calculated by integrating the frequency-integrated 
    intensities within the scaled ellipses,
    \begin{equation}
        L_\mathrm{clump} = 
        4\pi d^2\Omega_\mathrm{pix}\sum{I_\mathrm{int}}.
    \end{equation}

    The resulting dust temperatures, column densities, number densities,
    mass surface densities, 
    masses and luminosities
    are given in columns 3-8 of Table \ref{tb-para}, and the statistics
    are presented in Table \ref{tb-stats}. The candidate HMSCs are in general,
    cold and dense with a median temperature of 16 K and a median column density of 
    $4.4\times10^{22}$ cm$^{-2}$. The 
    masses range  between 8 and $6.1\times10^4$ \msun, with a median value of 1019 \msun~and 
    a mean of 3384 \msun. The luminosities vary 
    from 9 to $1.4\times10^5$ \lsun, with the mean and median
    luminosities being 1329 and 6838 \lsun, respectively.

    \begin{figure}
    	\centering
    	\includegraphics[width=0.46\textwidth]{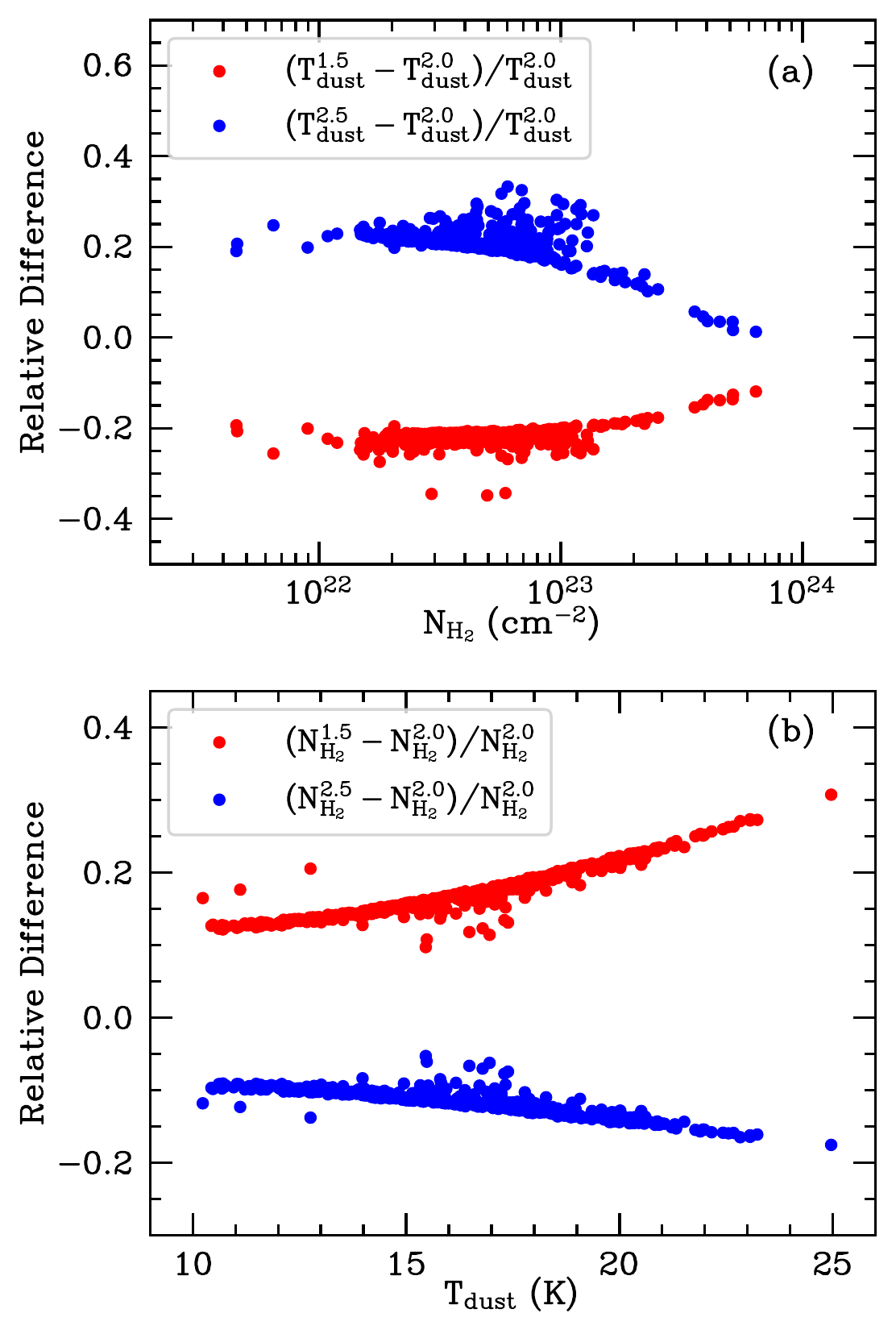}
    	\caption{Relative differences of column densities and 
    		dust temperatures when using different dust
    		emissivity indices (red symbols for $\beta = 1.5$ and
    		blue ones for $\beta = 2.5$).
    	}\label{fig:beta-effect}
    \end{figure}
    
    The uncertainties of the parameters largely originate from the uncertainties
    in distance and dust properties. The typical distance uncertainty is
    about 10\% and can propagates to other parameters. The precision of
    many parameters would heavily depend on the dust opacity which 
    is subject to a factor of 2 uncertainty \citep{1994A&A...291..943O}.
    The dust emissivity index can also largely influence the parameters 
    from SED fits. Shown in Figure \ref{fig:beta-effect} are the 
    relative changes of column densities and 
    dust temperatures at the peak positions of all of the HMSCs 
	when using different dust emissivity indices. The dust temperatures would increase by 
    $13\%-28\%$ if an index of 1.5 is adopted, and 
    decrease by $5\%-18\%$ if 2.5 is used. An index of 1.5
    would lead the column densities to decrease by $10\%-35\%$, while
    an index of 2.5 can enlarge the column densities by $2\%-35\%$.

	\begin{deluxetable*}{lccccccccc}
		\tablecolumns{10} 
		\tablewidth{0pt}
		\tablecaption{Statistics of some parameters for clumps at different stages \label{tb-stats}}
		\tablehead{
			\colhead{ } &\colhead{Distance}& \colhead{$r_\mathrm{eq}$}& 
			\colhead{$T_\mathrm{dust}$}& \colhead{$N_\mathrm{H_2}$}& 
			\colhead{$n_\mathrm{H_2}$}& 
			\colhead{$\Sigma_\mathrm{mass}$}&  \colhead{$M_\mathrm{cl}$}& 
			\colhead{$L_\mathrm{cl}$}& \colhead{$L_\mathrm{cl}/M_\mathrm{cl}$}\\
			\colhead{} & \colhead{(kpc)}  & \colhead{(pc)}  & \colhead{(K)}  & 
			\colhead{($10^{22}$ cm$^{-2}$)}  & \colhead{($10^{4}$ cm$^{-3}$)}  & 
			\colhead{(g cm$^{-2}$)}  & \colhead{($M_\odot$)}  & 
			\colhead{($L_\odot$)} & \colhead{($L_\odot/M_\odot$)} 
		}
		\startdata
		\multicolumn{10}{l}{Starless clumps} \\
		\midrule
		Min &          0.3 &       0.05 &      10.23 &       0.45 &       0.08 &       0.02 &    7.8e+00 &    8.6e+00 &       0.09 \\
		Max &         18.3 &       3.57 &      24.96 &      64.07 &     236.42 &       2.14 &    6.1e+04 &    1.4e+05 &      22.12 \\
		Median &          6.0 &       0.65 &      16.17 &       4.37 &       1.17 &       0.15 &    1.0e+03 &    1.3e+03 &       1.60 \\
		Mean &          7.1 &       0.90 &      16.22 &       5.90 &       3.48 &       0.22 &    3.4e+03 &    6.8e+03 &       2.54 \\
		\midrule
		\multicolumn{10}{l}{Clumps associated with methanol masers} \\
		\midrule
		Min &          0.7 &       0.07 &      11.88 &       0.71 &       0.05 &       0.03 &    6.3e+00 &    3.5e+01 &       0.21 \\
		Max &         22.0 &       3.05 &      47.09 &     342.80 &     237.24 &      31.34 &    2.9e+05 &    2.9e+06 &     400.20 \\
		Median &          8.1 &       0.67 &      20.31 &       5.27 &       2.08 &       0.25 &    1.6e+03 &    8.7e+03 &       5.31 \\
		Mean &          8.0 &       0.75 &      20.81 &       9.68 &       5.36 &       0.49 &    5.0e+03 &    3.6e+04 &       9.48 \\
		\midrule
		\multicolumn{10}{l}{Clumps associated with \hii~regions} \\
		\midrule
	    Min &          0.6 &       0.07 &      14.29 &       0.16 &       0.06 &       0.01 &    5.8e+00 &    7.3e+01 &       0.84 \\
	    Max &         20.9 &       2.81 &      51.87 &      43.71 &      51.95 &       3.19 &    2.5e+04 &    4.2e+06 &     524.46 \\
	    Median &          8.0 &       0.71 &      22.15 &       4.55 &       1.27 &       0.16 &    1.2e+03 &    1.4e+04 &       9.52 \\
	    Mean &          8.1 &       0.81 &      22.97 &       6.36 &       3.11 &       0.28 &    2.5e+03 &    5.4e+04 &      17.47  
		\enddata
		%\tablecomments{This table is available in its entirety in a machine-readable form in the online journal. A portion is shown here for guidance regarding its form and content.}
%		\tablenotetext{}{\textit{Note ---} All the listed parameters, 
%			except for $T_\mathrm{dust}$ and $N_\mathrm{H_2}$, are for the sample
%			with distances.}
	\end{deluxetable*}

\section{Discussion} \label{sec:discuss}

   \begin{figure}
	\centering
	\includegraphics[width=0.4\textwidth]{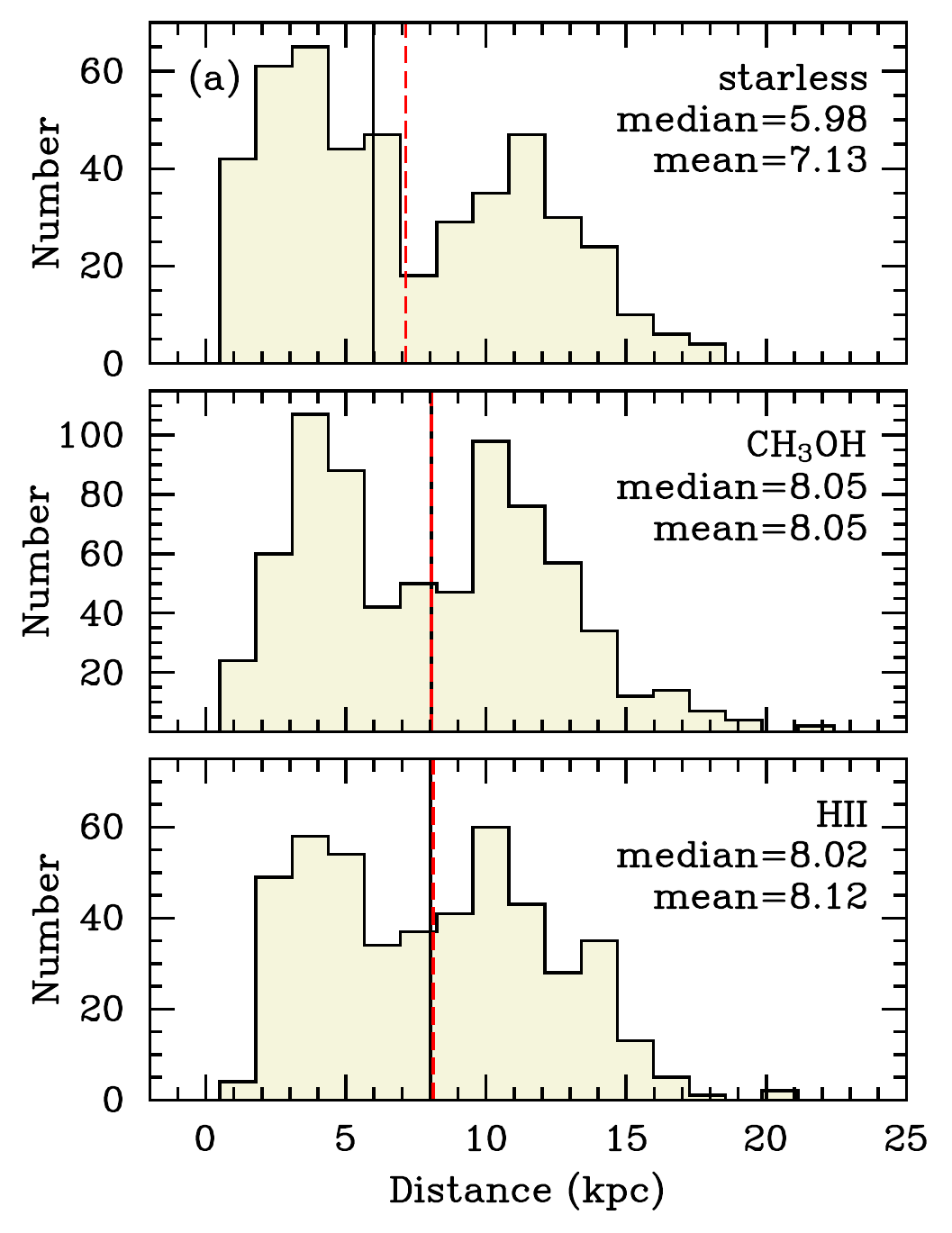}
	\includegraphics[width=0.4\textwidth]{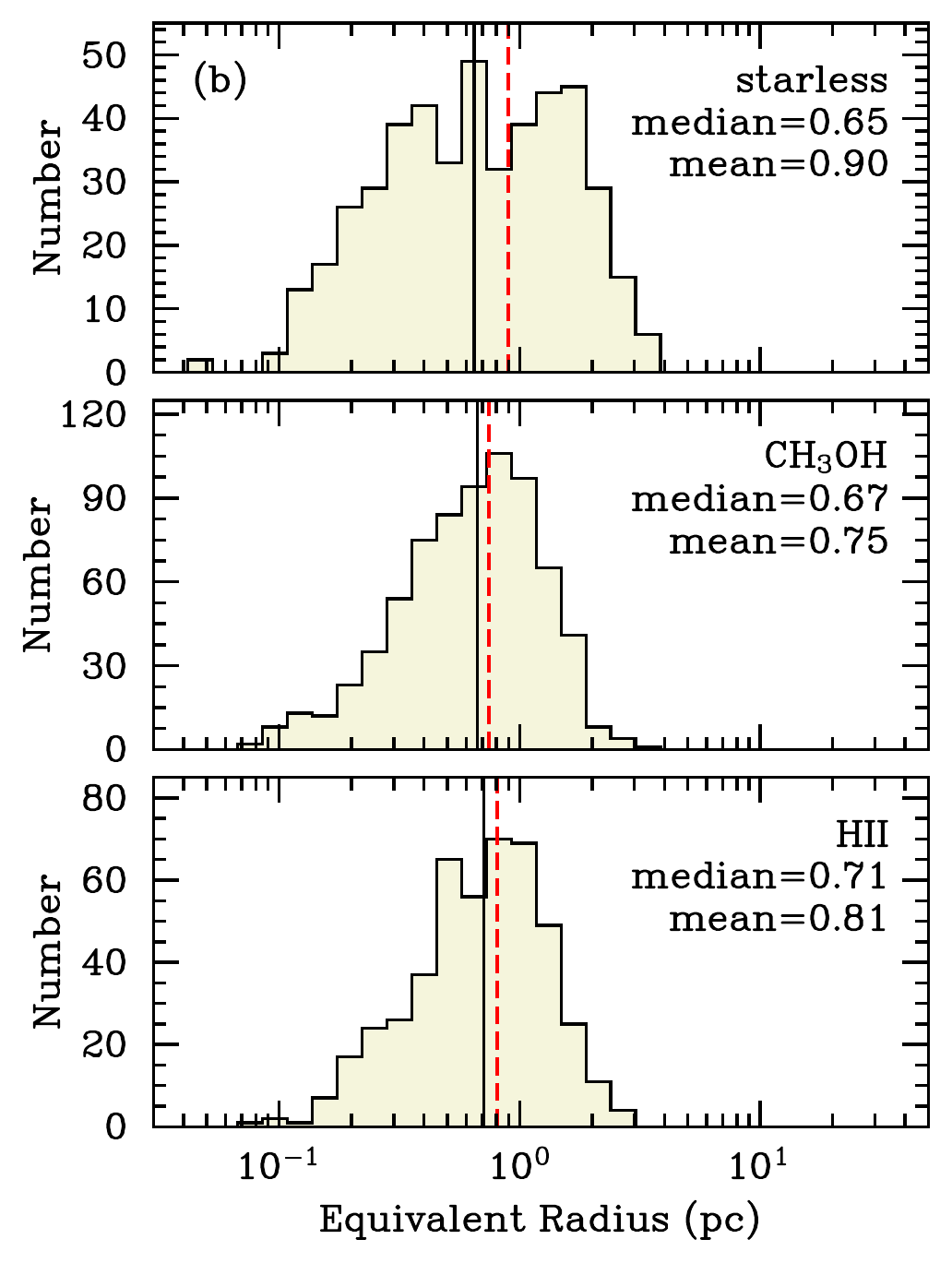}
	\caption{Histograms of distance (a) and equivalent radius (b, 
		see Section \ref{distance})
		of HMSC candidates, clumps associated with
		methanol masers and \hii~regions. 
		The black solid and red dashed vertical 
		lines mark the median and mean values.}\label{fig:histDist}
\end{figure}

\begin{figure*}
	\centering
	\includegraphics[width=0.33\textwidth]{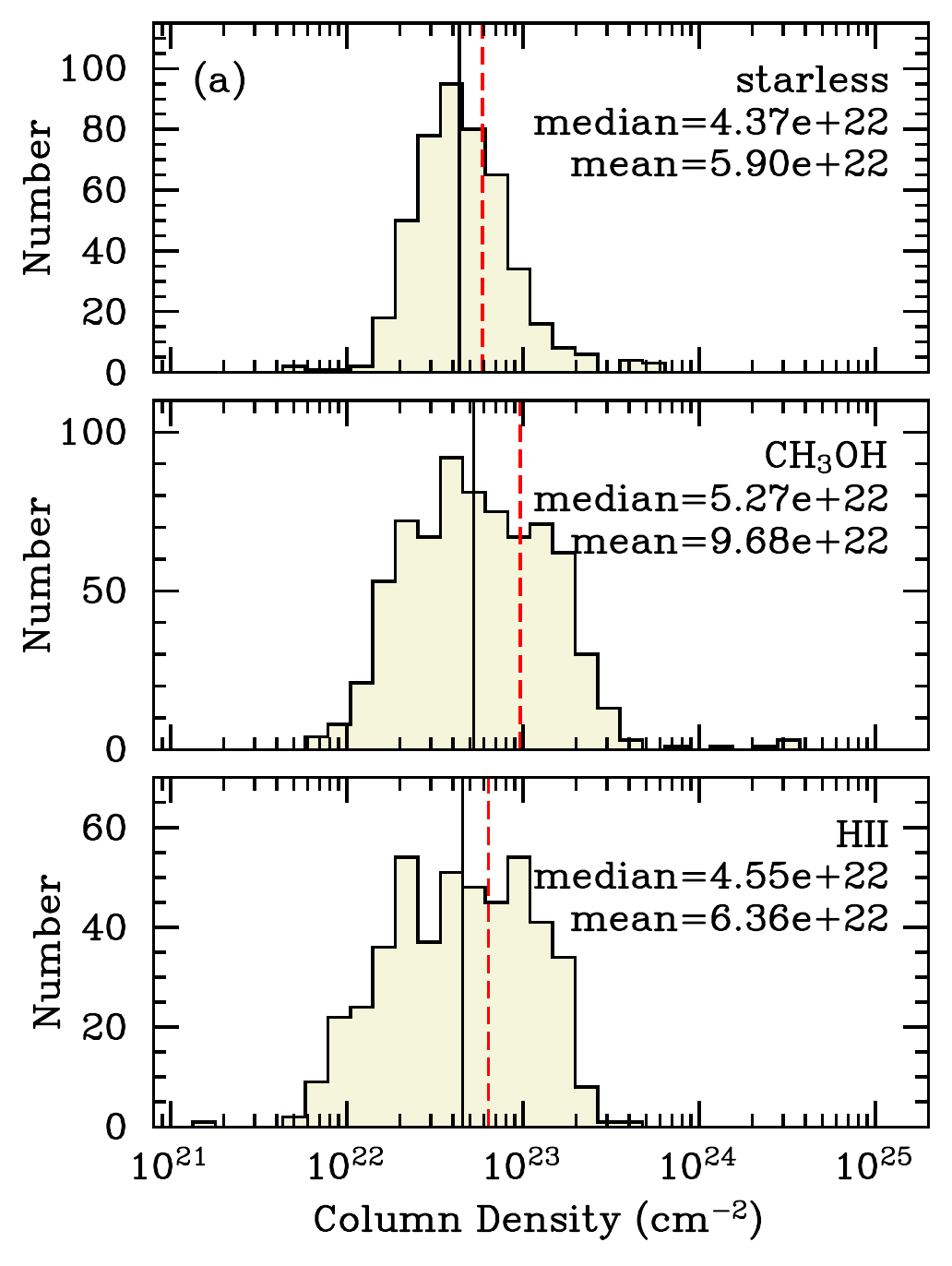}
	\includegraphics[width=0.33\textwidth]{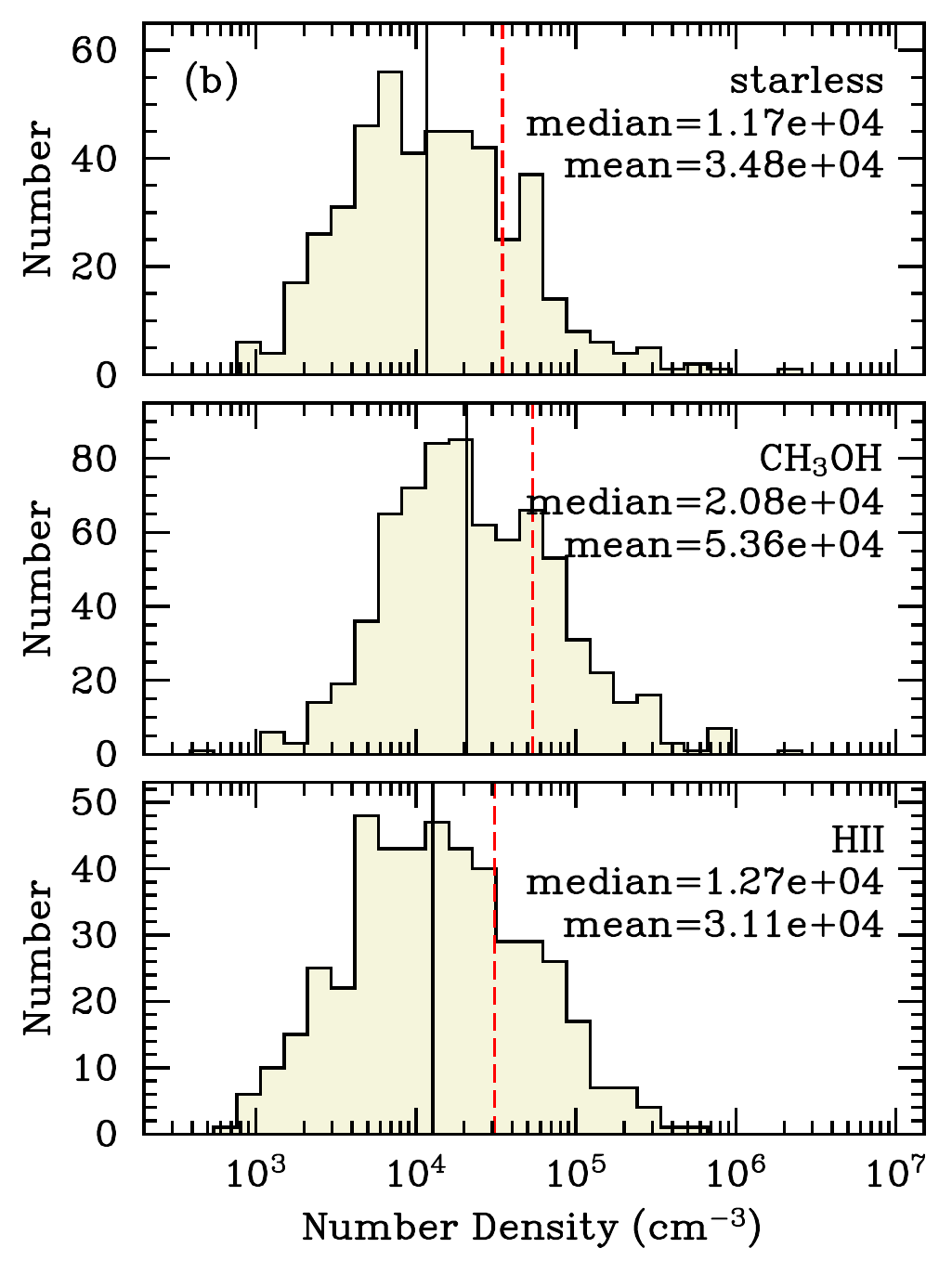}
	\includegraphics[width=0.33\textwidth]{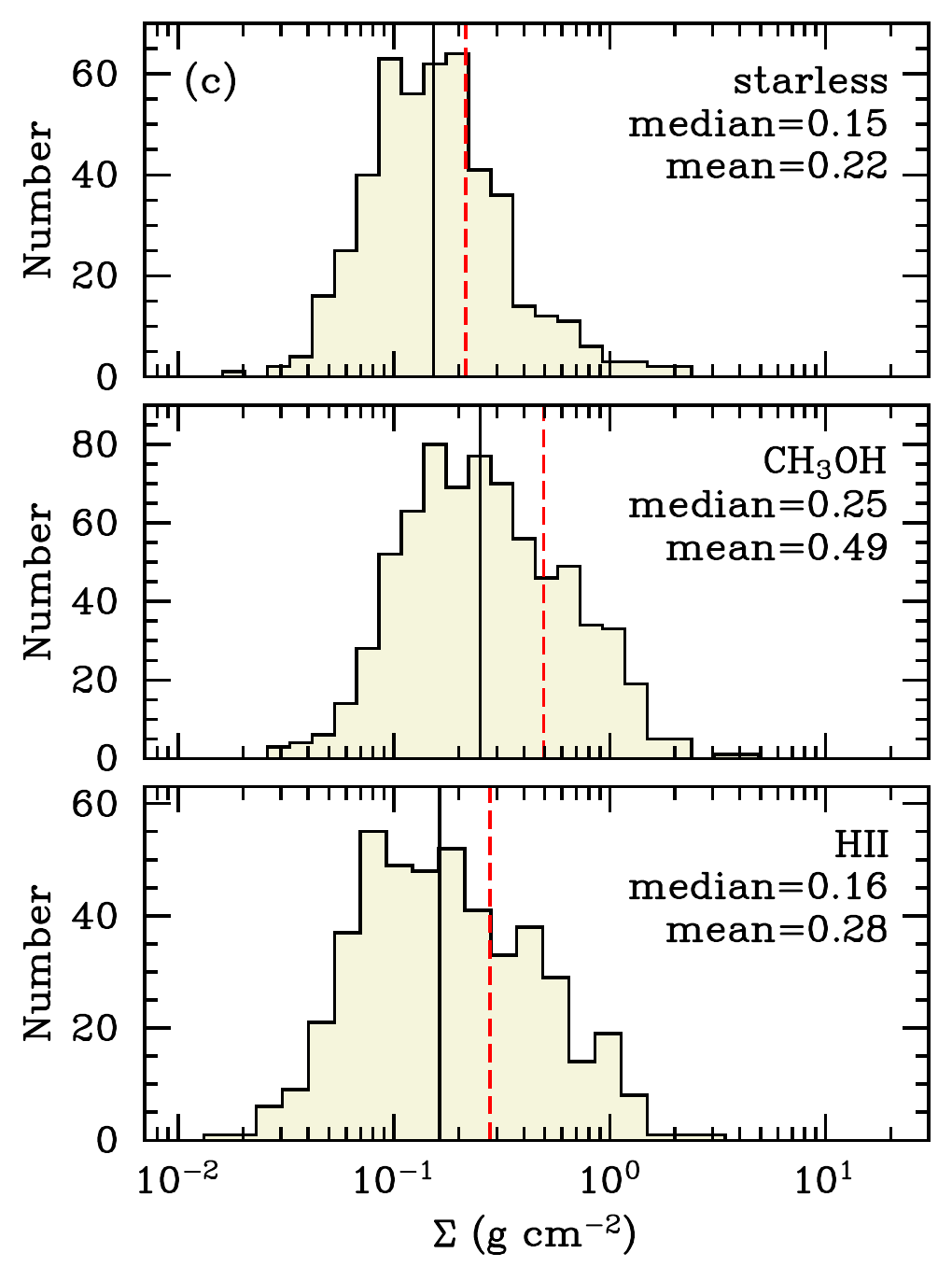}
	\caption{Histograms of H$_2$ column density (a)
		H$_2$ number density (b) and mass surface density (c)
		of HMSC candidates, clumps associated with
		methanol masers and \hii~regions. 
		The black solid and red dashed vertical 
		lines mark the median and mean values, respectively.}\label{fig:density}
\end{figure*}

\begin{figure*}
	\centering
	\includegraphics[width=0.33\textwidth]{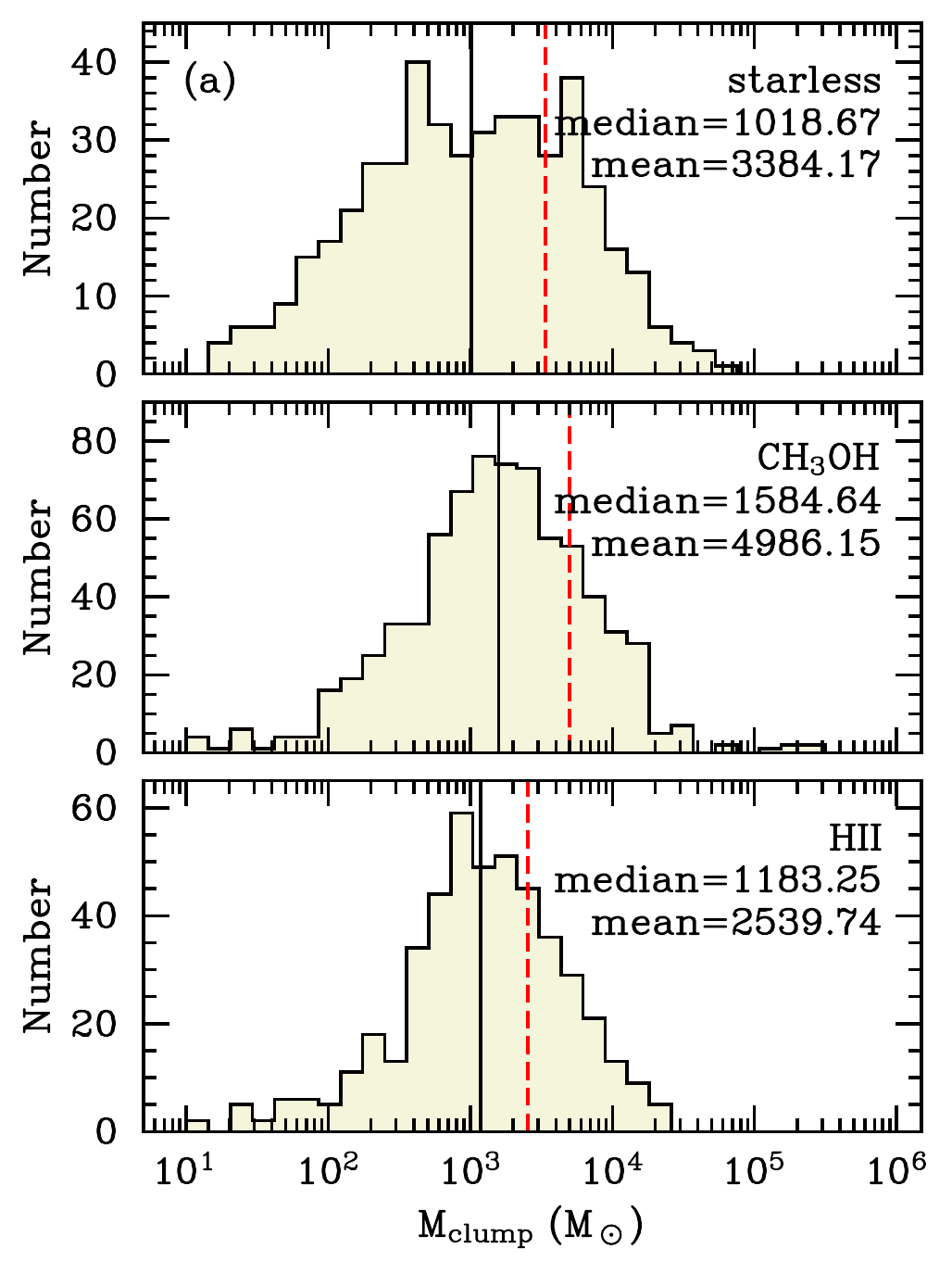}
	\includegraphics[width=0.33\textwidth]{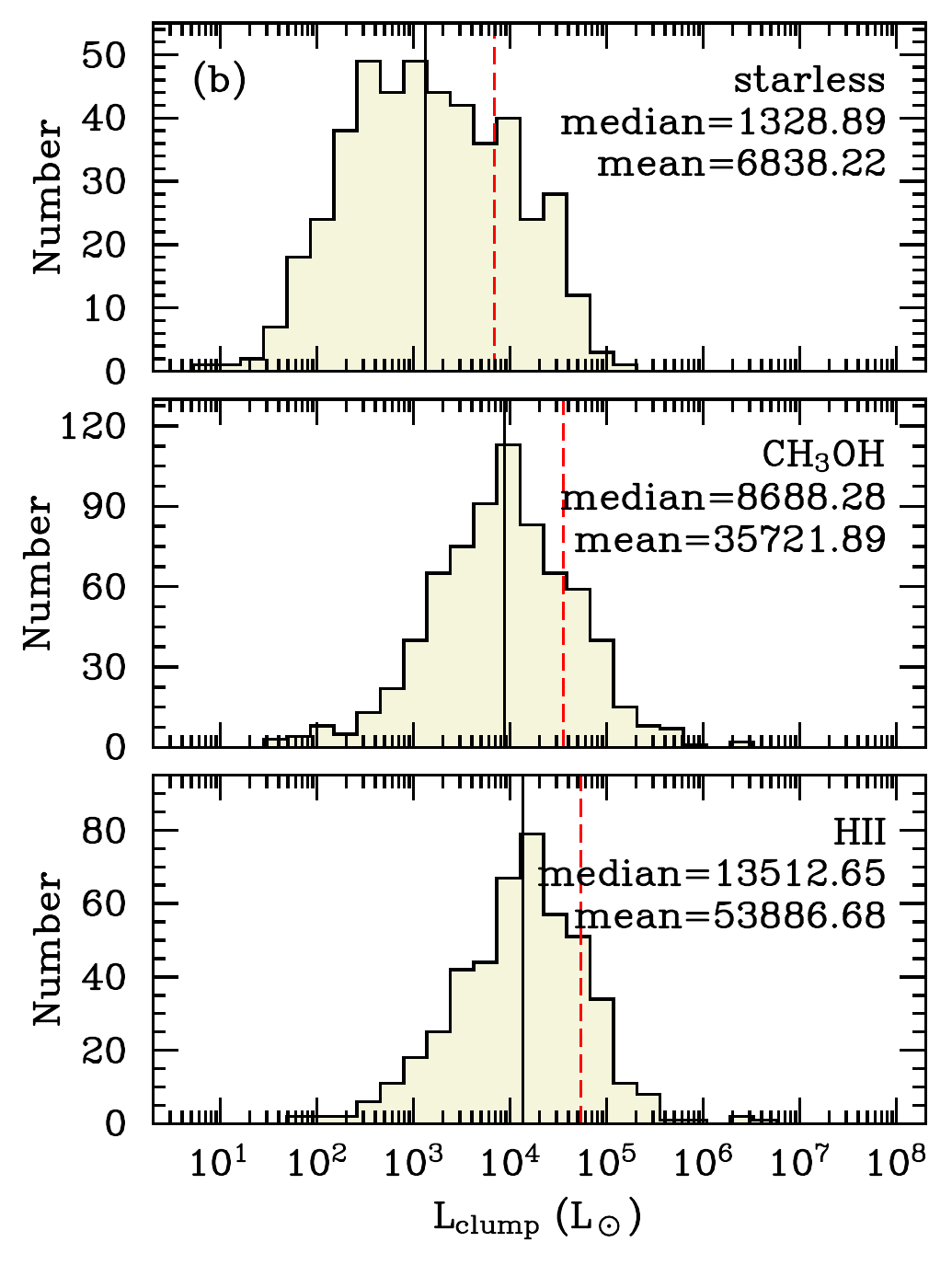}
	\includegraphics[width=0.33\textwidth]{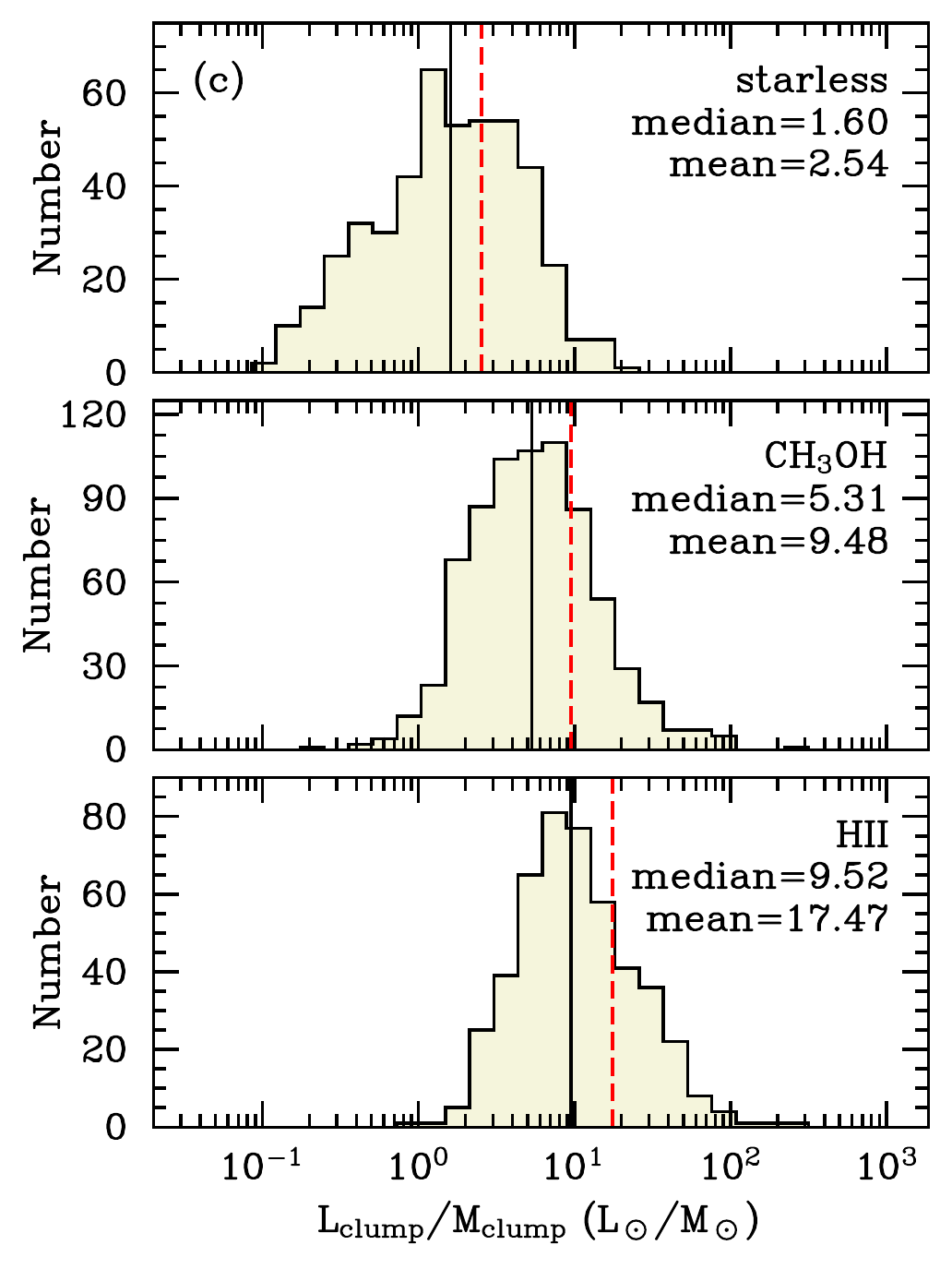}
	\caption{Histograms of clump mass (a), clump luminosity (b)
		and  luminosity-to-mass ratio (c)
		of HMSC candidates, clumps associated with
		methanol masers and \hii~regions. 
		The black solid and red dashed vertical 
		lines mark the median and mean values,respectively.}\label{fig:histMass}
\end{figure*}

	In order to investigate whether the HMSC candidates 
	genuinely represent an early phase of high-mass star formation, it is desirable to
	compare their properties with those of other samples of young high-mass star formation
	regions.  We have obtained dust properties for 728 clumps associated with 6.7 GHz Class II methanol
	masers (methanol clumps, hereafter) and 469  
	\hii~regions (\hii~clumps, hereafter) following the procedure given 
	in Section \ref{sect:dust}. The methanol clumps were selected 
	based on a spatial cross-match between ATLASGAL sources and 6.7 GHz 
	methanol masers from the MMB survey 
	\citep{2010MNRAS.404.1029C,2011MNRAS.417.1964C,2010MNRAS.409..913G,
		2012MNRAS.420.3108G,2015MNRAS.450.4109B}. Only sources residing in the
	inner Galactic plane were included in the sample. The sources associated with 
	\hii~regions were taken from \citet{2014MNRAS.443.1555U}. Similarly, only sources
	with $|l|<60\deg$ and $|b|<1\deg$ were considered. To maintain consistency with the
	HMSC candidate sample, we only included sources 
	with peak intensities $ >0.5 $ Jy beam$^{-1}$ at 870 \micron. 
	In the following sections, we show that the HMSC candidates are entities similar to 
	the clumps associated with methanol maser and \hii~region clumps, which are known
	to have formed high-mass stars, but the HMSC clumps are at an earlier evolutionary stage.
	
	\subsection{Comparison of Clump Properties}
	
	The starless, methanol and \hii~samples have {463 (100\%)}, 
	722 (99\%), and 464 (99\%) sources 
	that have systemic velocity information, enabling us to 
	obtain distance measurements using the parallax-based distance
	estimator (see Section \ref{sec:distance}).
	Histograms of the distances for the three groups of sources are 
	shown in Figure \ref{fig:histDist} (a). The methanol group has similar 
	median and mean distances to the \hii~group, while starless clumps
	tend to be nearer, with smaller median and mean distances.
	The fraction of sources at distances farther than 8 kpc are 41\%, 50\%, and 50\% for
	the starless, methanol and \hii~clumps, respectively. 
	The greater mean distance in the star-forming groups
	is consistent with the fact that the more evolved 
	clumps tend to show stronger emission at 870~\micron~\citep{2016MNRAS.461.2288H},
	making the detection of methanol and~\hii~clumps at 
	the far distance easier than for the starless sample. 
	
	The difference in the distance distribution of the samples raises the question as to whether they
	are truly comparable groups or if they represent overlapping, but different populations?
	Figures \ref{fig:histDist} (b), \ref{fig:density} (a-c), 
	and \ref{fig:histMass} (a), show that the three samples have very similar
	distributions of sizes, densities, and masses, suggesting they
	are comparable groups, but at different stages of the star formation process
	(see subsection \ref{sec:stages}). In contrast, increasing 
	densities and masses from starless to \hii~clumps have been
	reported by \citet{2016MNRAS.461.2288H} and \citet{2016ApJ...822...59S}.
	Compared to the clumps investigated in \citet{2016MNRAS.461.2288H} and 
	\citet{,2016ApJ...822...59S}, those investigated in this work are
	in general higher mass, as we set a minimum 870 \micron~
	peak intensity of 0.5 Jy beam$^{-1}$. We suggest that the inclusion of some lower-mass sources
	in \citet{2016MNRAS.461.2288H} and \citet{2016ApJ...822...59S},
	especially for objects at early stages, is the main reason 
	that they observe increasing masses and densities from starless
	to \hii~clumps.
%	However, data at only one band were used to estimate 
%	column densities and masses in \citet{2016MNRAS.461.2288H} (870~\micron) 
%	and \citet{2016ApJ...822...59S} (1.1 mm). The difference between tendencies
%	in this work and the literature could be due to different methods used for 
%	estimating densities and masses.

	\subsection{High-mass Star Birth-sites}

	\begin{figure}
		\centering
		\includegraphics[width=0.48\textwidth]{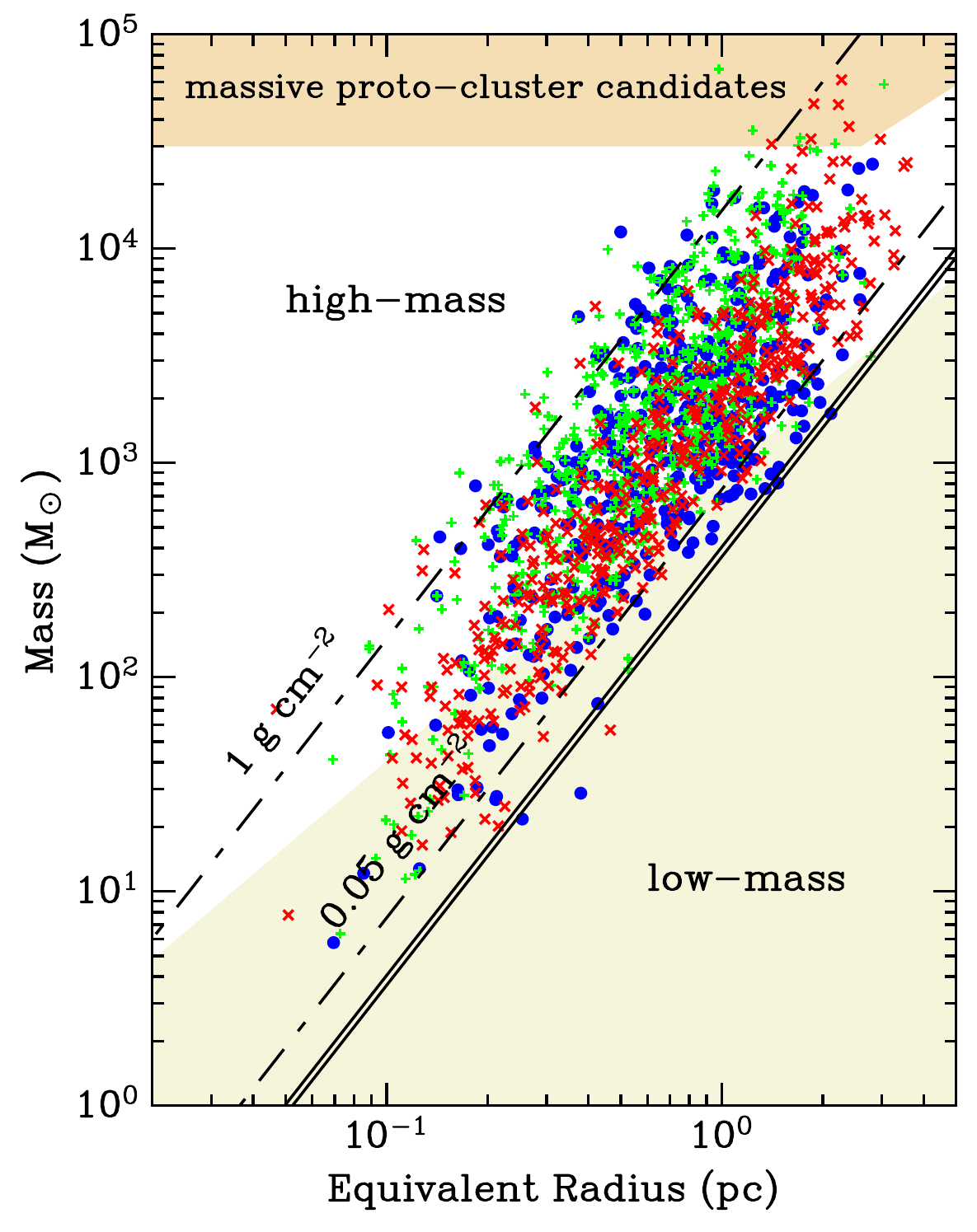}
		\caption{Clump mass as a function of equivalent radius 
			(see Section \ref{distance}) for 
			HMSC candidates (red crosses), 
			clumps associated with methanol masers (green plus symbols)
			and HII regions (blue filled circles). The unshaded area delimits
			high-mass star formation regions. The threshold, 
			$M>870~M_\odot~(r/\mathrm{pc})^{1.33}$, is adopted from
			\citet{2010ApJ...723L...7K}. The mass surface 
			density thresholds for ``efficient star formation''
			of 116 $M_\odot$ pc$^{-2}$ ($\sim0.024$ g cm$^{-2}$) 
			from \citet{2010ApJ...724..687L}
			and 129  $M_\odot$ pc$^{-2}$ ($\sim0.027$ g cm$^{-2}$) from \citet{2010ApJ...723.1019H}
			are shown as black solid lines. The upper and lower 
			dash-dotted lines give two mass surface density cuts of 
			0.05 g cm$^{-2}$ and 1 g cm$^{-2}$. 
			The upper shaded region indicates the parameter space
			for massive proto-clusters defined in \citet{2012ApJ...758L..28B}.}\label{fig:MvsR}
		\vskip 0.5cm
	\end{figure}

	To establish whether we have assembled a robust sample of HMSC candidates, it is crucial to assess 
	their potential to form high-mass stars. 
	Shown in Figure \ref{fig:MvsR} is the mass-radius diagram on which the 
	 starless, methanol, and \hii~clumps
	are presented as red crosses, green plus symbols, and filled blue
	circles, respectively. Also overplotted are several thresholds for 
	star formation determined from local clouds. All of the sources with the
	exception of one starless, three methanol and three \hii~clumps meet the criteria for
	``efficient'' star formation of $116$ \msun~pc$^{-2}$ ($\sim0.024$ g cm$^{-2}$) and 
	$129$ \msun~pc$^{-2}$ ($\sim0.027$ g cm$^{-2}$,
	 solid lines) from 
	\citet{2010ApJ...724..687L} and \citet{2010ApJ...723.1019H}. 
	
	A more restrictive high-mass star-formation
	criterion of $M\geq870~M_\odot\ (R_\mathrm{eq}/\mathrm{pc})^{1.33}$ 
	was suggested by \citet{2010ApJ...723L...7K} from observations of
	nearby clouds. About 82\% (378/463) HMSC candidates
	are above this threshold, and are potential high-mass
	star forming regions. The fractions for the methanol and \hii~groups
	fulfilling this threshold are 90\% and 80\%, respectively. 
	
%	Using the stellar IMF
%	from \citet{2001MNRAS.322..231K} and assuming a typical 
%	star formation efficiency of 0.3, \citet{2016ApJ...822...59S}
%	proposed a minimum mass of 320 \msun~for a clump to form
%	at least one massive star ($>8$ \msun). Among the 319 HMSC candidates
%	with distance estimates, there are 242 ($\sim69\%$) clumps
%	having masses above this threshold. Note that the required minimum 
%	mass for producing high-mass stars is sensitive to the assumed star 
%	formation efficiency. Using a higher value of 0.5, this threshold 
%	would decrease to about 190 \msun, and the fraction of clumps above this 
%	threshold in our sample would increase to about 80\%.
	
	Mass surface density, $\Sigma_\mathrm{mass}$, is another commonly used parameter to 
	assess the high-mass star formation potential. 
	\citet{2008Natur.451.1082K} suggest that a minimum mass surface density 
	of 1 g cm$^{-2}$  is required to prevent fragmentation into 
	low-mass cores through radiative feedback, thus
	allowing high-mass star formation.  However, this threshold
	is relatively uncertain, and magnetic fields, which can help 
	prevent fragmentation, were not considered in the calculations. 
	High-mass clumps and cores with $0.05\le\Sigma\le0.5$ g cm$^{-2}$ 
	are indeed reported in the literature 
	\citep{2012ApJ...754....5B,2013A&A...555A.112P,2013ApJ...779...96T}. 
	In a recent study based on ATLASGAL clumps, \citet{2014MNRAS.443.1555U}
	suggested a less stringent empirical threshold of 0.05 g cm$^{-2}$.
	If we adopt a threshold of 1 g cm$^{-2}$, the fraction of high-mass
	star forming clump candidates which exceed this is less than 10\%, even for the methanol and 
	\hii~groups where high-mass star formation is known to be occurring. In contrast, more than 90\%
	of clumps exceed the less stringent 0.05 g cm$^{-2}$ threshold. Thus, 
	it appears that the \citet{2014MNRAS.443.1555U} threshold is more useful for observations
	that average over a clump. Among the
	HMSC candidates with distances, there are 448 clumps ($\sim97\%$) fulfilling
	this threshold and hence having the potential to form high-mass stars.
	
%	There are only two HMSC candidates 
%	(G012.9459-0.2488 and G354.9437-0.5381) with 
%	masses $>3\times10^4$ \msun, which locates them in 
%	the parameter space for massive proto-clusters (PMCs) (upper 
%	shaped region in Figure \ref{fig:MvsR})
%	defined by \citet{2012ApJ...758L..28B}. 

	\subsection{The Very Early Phases of Star Formation} \label{sec:stages}
	
	\begin{figure}
		\centering
		\includegraphics[width=0.4\textwidth]{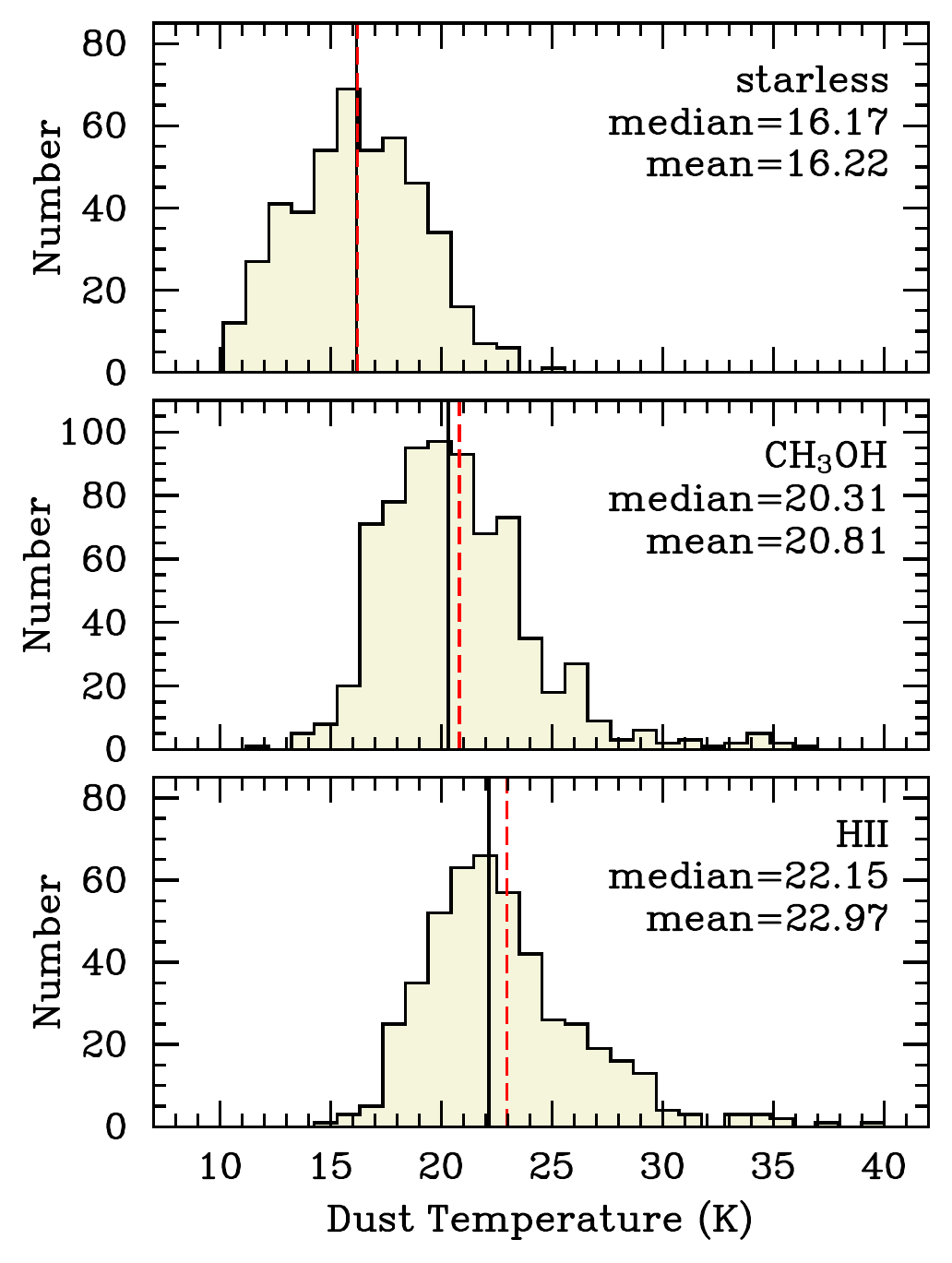}
		\caption{Histograms of dust temperature 
			of HMSC candidates (a), clumps associated with
			methanol masers (b)  and \hii~regions (c). 
			The black solid and red dashed vertical 
			lines mark the median and mean values. 
		}\label{fig:histTdust}
	\end{figure}

	\begin{figure}
		\centering
		\includegraphics[width=0.48\textwidth]{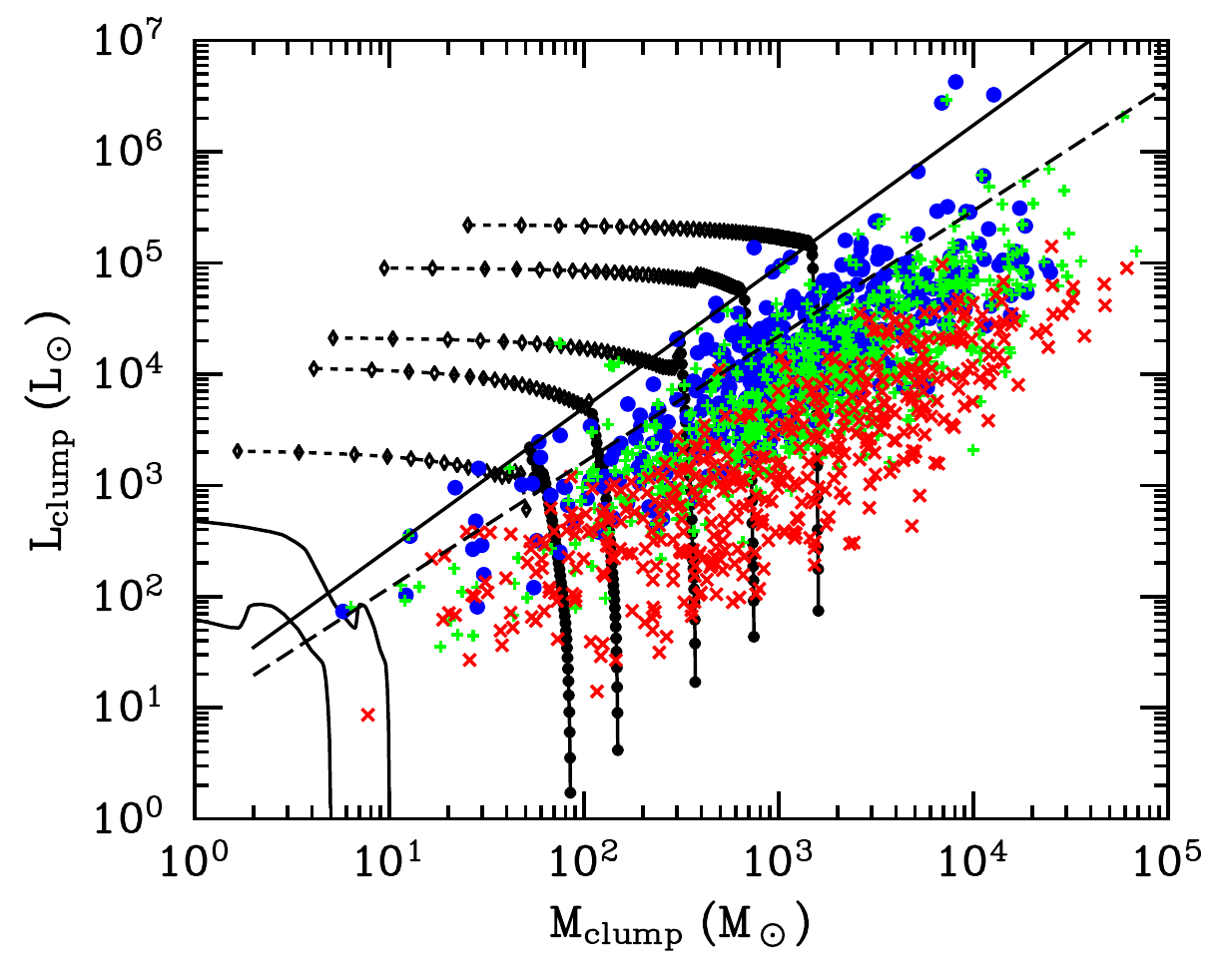}
		\caption{Luminosity-mass diagram for 
			HMSC candidates (red crosses), 
			clumps associated with methanol masers (green plus symbols)
			and HII regions (blue filled circles). 
			Evolution tracks for stars with final masses
			of 2.0, 4.0, 6.5, 8.0, 13.5, 18.0, and 35.0 \msun~
			are from \citet[][solid tracks]{1996A&A...309..827S} and
			\citet[][tracks with symbols]{2008A&A...481..345M}. The solid and dashed
			lines are the best log-log fit for Class I and Class 0 sources 
			extrapolated in the high-mass regime by \citet{2008A&A...481..345M}.
		}\label{fig:LvsM}
		\vspace{0.2cm}
	\end{figure}
	
	\begin{figure}
		\centering
		\includegraphics[width=0.45\textwidth]{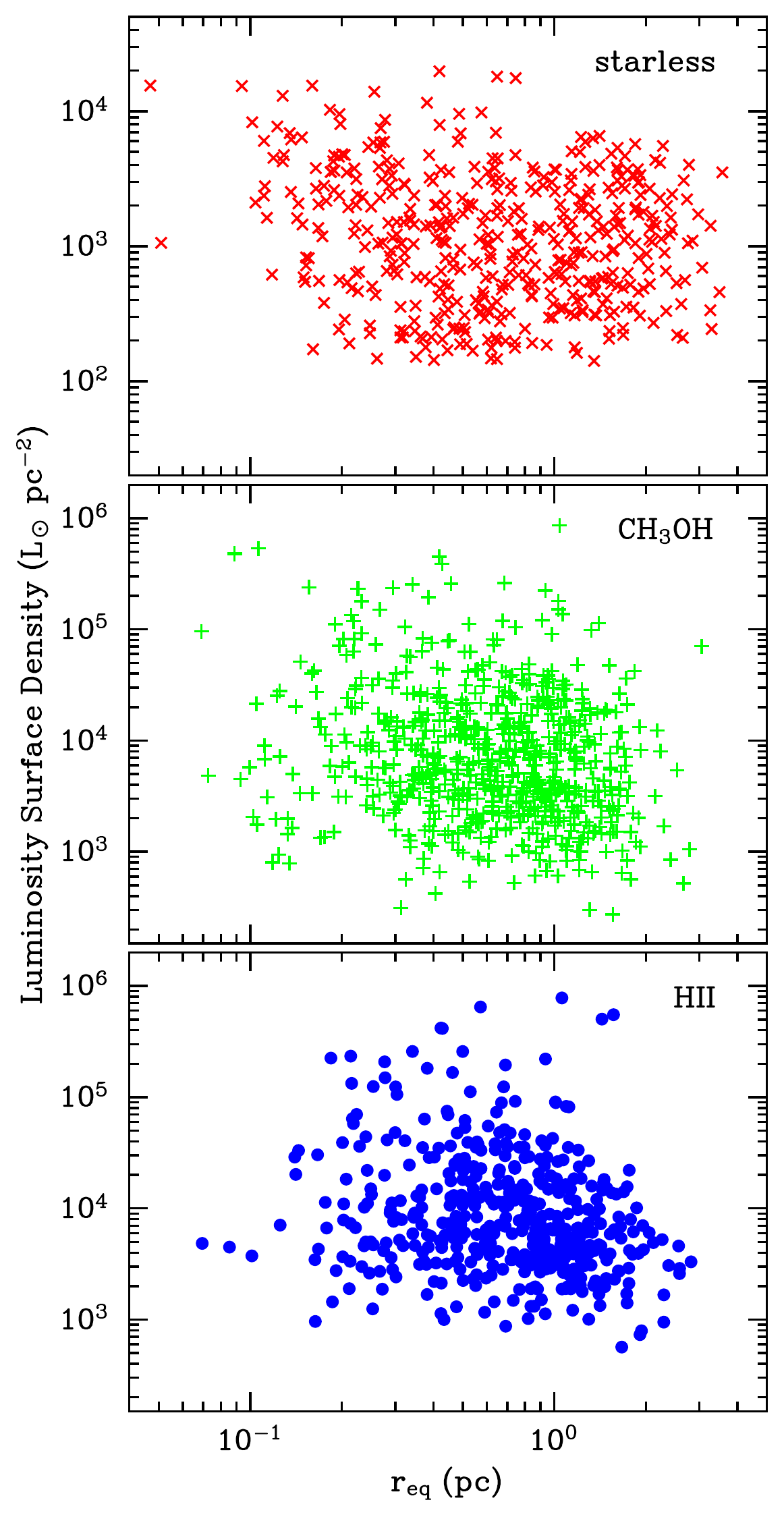}
		\caption{Luminosity surface density versus equivalent radius
			(see Section \ref{distance}) for 
			HMSC candidates (upper panel), 
			clumps associated with methanol masers (middle panel)
			and HII regions (lower panel). Note that the 
			ordinate in the top panel covers 
			lower values than in the two 
			lower panels.
		}\label{fig:LumSurfaceDense}
		\vspace{0.2cm}
	\end{figure}
	
	As a clump evolves from a quiescent phase to one with active 
	star formation, radiative heating from star formation is 
	expected to raise the dust temperature. As \hii~regions
	are more evolved indicators of star formation than
	methanol masers, \hii~clumps should be subject to stronger
	radiative heating. 
	Thus, dust temperature, in general, should
	serve as a tracer of evolutionary stage \citep{2002ApJS..143..469M}. 
	Figure \ref{fig:histTdust} shows the distribution of dust 
	temperatures for the three groups. The median values 
	are 16.2, 20.3, and 22.2 for starless, methanol, and \hii~
	clumps, respectively. This trend of increasing 
	dust temperature is consistent with an evolutionary sequence 
	from starless to \hii~clumps. 
	A K-S test shows that the probability
	for the three distributions to be the same is smaller than 0.1\%,
	indicating that the HMSC candidates represent an 
	earlier phase of star formation.
	
	The distribution of luminosities for the three samples are presented 
	in Figure \ref{fig:histMass}(b), where
	again, an increasing trend can be observed. The median luminosities 
	are 1329, 8688, and 13513 \lsun~for starless, methanol, 
	and \hii~clumps, respectively. We attribute this difference in luminosity
	to emission arising from warm cores	with embedded protostars. In starless clumps, emission from 
	dust envelopes, heated externally, dominates the luminosity.  In contrast, warmer
	cores in methanol and \hii~clumps significantly contribute 
	to the luminosities. Since there can be significant
	emission at wavelengths shorter than 70 \micron, the genuine
	luminosities for clumps with embedded warm cores will be even higher than
	the values estimated in this work. 
	
	Another effective tool for diagnosing different stages of dense structures 
	in molecular clouds is the luminosity--mass
	 ($L_\mathrm{clump}-M_\mathrm{clump}$) diagram, 
	on which sources at different phases of star formation
	can be readily distinguished \citep{2008A&A...481..345M,2016ApJ...826L...8M}.
	This diagram for high-mass star formation was introduced
	by \citet{2008A&A...481..345M} based on the two-phase
	model of \citet{2003ApJ...585..850M}. In the first phase,
	the mass of a core slightly decreases due to
	accretion and molecular outflows, while the luminosity increases
	significantly, and the source moves following an almost
	vertical track in the $L_\mathrm{clump}-M_\mathrm{clump}$ diagram. 
	In the second phase, the surrounding material is expelled through 
	radiation and molecular outflows. With a nearly constant luminosity,
	the object follows an almost horizontal path. Although the 
	evolutionary tracks have been initially modeled for single cores, 
	the $L_\mathrm{clump}-M_\mathrm{clump}$ diagram has 
	been also frequently used to discuss the evolution of clumps 
	\citep[e.g.,][]{2010A&A...518L..97E,2013ApJ...772...45E,
		2015MNRAS.451.3089T,2016A&A...585A.149W}
	
	The $L_\mathrm{clump}-M_\mathrm{clump}$ diagram is shown in 
	Figure \ref{fig:LvsM} with the same symbol convention as that 
	in Figure \ref{fig:MvsR}.  Also overplotted are the theoretical 
	evolutionary tracks for the low- and high-mass regimes adopted
	from \citet{2008A&A...481..345M}. The best log-log fits for Class I
	and Class 0 sources extrapolated in the high-mass regime 
	by \citet{2008A&A...481..345M} are shown as solid and 
	dashed lines. Although there is a degree of overlap, segregated parameter
	spaces are occupied by different groups of clumps. With 
	a median luminosity to mass ratio ($L_\mathrm{clump}/M_\mathrm{clump}$) 
	about 1.6 $L_\odot/M_\odot$, the HMSC candidates 
	are mainly located towards the bottom-right of the diagram.  
	The $L_\mathrm{clump}/M_\mathrm{clump}$ values we find are comparable to those of starless clumps
	reported in \citet{2015MNRAS.451.3089T} and significantly 
	lower than those of known protostellar clumps 
	\citep[e.g.,][]{2014MNRAS.443.1555U,2015MNRAS.451.3089T}.
%	Compared to the dust opacities used to calculate the masses for cold clumps
%	in the literature
%	\citep[e.g.,][]{2010A&A...518L..97E,2013ApJ...772...45E,
%		2010ApJ...723L...7K,2015MNRAS.450.4043W}, 
%	the opacity adopted in this work
%	will produce estimates which are 30--40 percent higher.
%	If we had used the dust opacities same as in 
%	\citet{2010ApJ...723L...7K} and 
%	\citet{2010A&A...518L..97E,2013ApJ...772...45E}, 
%	the $L_\mathrm{clump}/M_\mathrm{clump}$
%	ratios would be scaled down by a factor of about 1.5,
%	leading to more than 60\% (196/319) HMSC candidates
%	fulfilling the criteria proposed by \cite{2016ApJ...826L...8M}.
	
	\citet{2016ApJ...826L...8M} suggest that 
	$L_\mathrm{clump}/M_\mathrm{clump}<1~L_\odot/M_\odot$ 
	is characteristic of starless clumps, however, 
	some of our HMSC candidates have a larger luminosity--mass ratio.
	A possible reason for many HMSC candidates having L/M ratios 
	higher than the suggested threshold of 1 \lsun/\msun~is that 
	some may be externally heated by the interstellar radiation field (ISRF).
	For a source externally heated by the ISRF, the luminosity surface density 
	would stay constant. In contrast, the luminosity 
	surface density for a star-forming clump will decrease 
	with increasing radius because of the internal heating from 
	embedded protostars, 
	which dominate the observed luminosities for smaller sources.
	Plots of luminosity surface density versus equivalent
	radius for the three samples are shown in Figure \ref{fig:LumSurfaceDense}. 
	Note that the ordinate in the top panel covers lower 
	values than in the two lower panels.
	A weak decreasing trend can be seen for methanol and \hii~
	clumps, while the HMSC candidates are scattered in 
	luminosity surface density versus radius parameter space.
	This is consistent with the methanol and \hii~clumps 
	being internally heated while the starless clumps are  
	externally heated by the ISRF. If this is the case, one 
	would expect smaller L/M ratios for starless clumps if the 
	masses and luminosities are measured over smaller regions. 
	When we calculate the luminosities and masses in just one 
	beam, the median, mean, and maximum 
	L/M ratios for the HMSC sample decrease from 1.60, 2.54, and 22.12 to 
	1.36, 2.19, and 20.59. In contrast, the L/M ratios
	increase by about 11\% and 30\% for methanol and 
	\hii~clumps. This provides further evidence of 
	external heating for the HMSC candidates and 
	internal heating for the the (methanol and \hii) star-forming clumps.
		
	\subsection{Comparison with the BGPS Starless Clumps}\label{sec-com-BGPS}
	
	\begin{figure}
		\centering
		\includegraphics[width=0.45\textwidth]{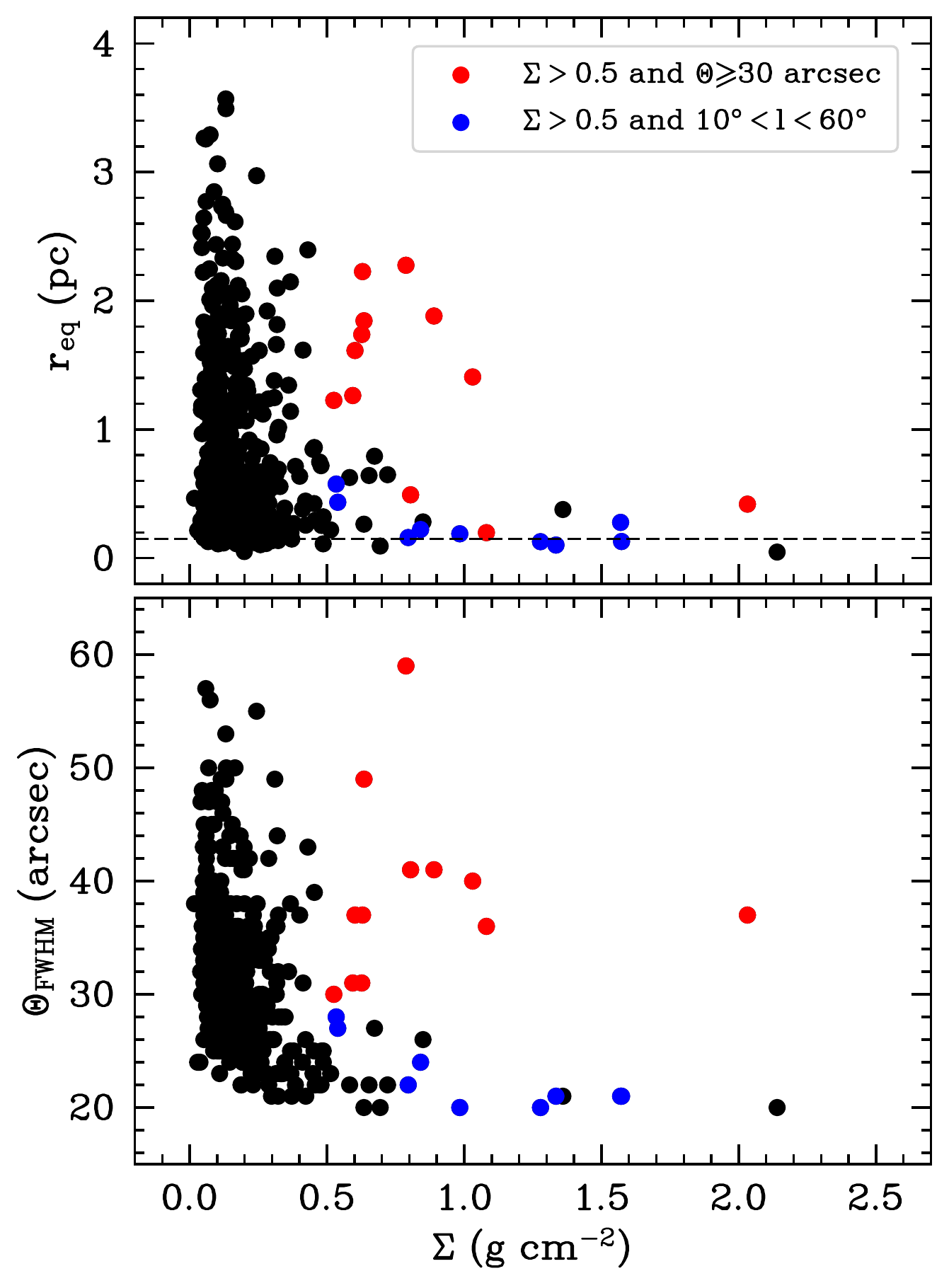}
		\caption{Equivalent radius (a, see Section \ref{distance}) and angular size (b)
			versus mass surface density. The sources with mass
			surface densities $>0.5$ g cm$^{-2}$ and angular sizes $>30$ 
			\arcsec~are shown in red, and the ones located in 
			$10\deg<l<60\deg$ in blue. The dashed line in the upper 
			panel marks a radius threshold of 0.15 pc for separating
			cores and clumps.
		}\label{fig:Size_Sigma}
		\vspace{0.2cm}
	\end{figure}

	\begin{deluxetable*}{lccccccccc}
	\tablecolumns{10} 
	\tablewidth{0pt}
	\tablecaption{Physical parameters for starless core candidates \label{tb-cores}}
	\tablehead{
		\colhead{ } &\colhead{Distance}& \colhead{$r_\mathrm{eq}$}& 
		\colhead{$T_\mathrm{dust}$}& \colhead{$N_\mathrm{H_2}$}& 
		\colhead{$n_\mathrm{H_2}$}& 
		\colhead{$\Sigma_\mathrm{mass}$}&  \colhead{$M_\mathrm{cl}$}& 
		\colhead{$L_\mathrm{cl}$}& \colhead{$L_\mathrm{cl}/M_\mathrm{cl}$}\\
		\colhead{} & \colhead{(kpc)}  & \colhead{(pc)}  & \colhead{(K)}  & 
		\colhead{($10^{22}$ cm$^{-2}$)}  & \colhead{($10^{4}$ cm$^{-3}$)}  & 
		\colhead{(g cm$^{-2}$)}  & \colhead{($M_\odot$)}  & 
		\colhead{($L_\odot$)} & \colhead{($L_\odot/M_\odot$)} 
	}
	\startdata
	 G000.5184-0.6127 &   3.3 &  0.13 & 18.86 &  1.59 &   11.46 &  0.30 &      81.32 &     392.85 &  4.83 \\
	G004.4076+0.0993 &   2.9 &  0.14 & 18.13 &  2.08 &   12.26 &  0.32 &      89.93 &     359.99 &  4.00 \\
	G010.2144-0.3051 &   3.1 &  0.13 & 16.64 &  6.58 &   51.83 &  1.28 &     313.76 &     666.97 &  2.13 \\
	G014.4876-0.1274 &   3.1 &  0.13 & 13.84 &  6.69 &   63.20 &  1.57 &     393.71 &     247.41 &  0.63 \\
	G015.2169-0.4267 &   1.9 &  0.11 & 15.44 &  2.74 &   12.72 &  0.28 &      53.76 &      65.55 &  1.22 \\
	G035.2006-0.7253 &   2.2 &  0.10 & 15.22 &  7.67 &   68.13 &  1.33 &     206.63 &     267.73 &  1.30 \\
	G327.2585-0.6051 &   2.9 &  0.15 & 18.52 &  2.63 &   13.03 &  0.37 &     122.21 &     439.37 &  3.60 \\
	G328.2075-0.5865 &   2.7 &  0.09 & 18.81 &  2.34 &   38.33 &  0.69 &      91.82 &     424.89 &  4.63 \\
	G350.7947+0.9075 &   1.4 &  0.12 & 15.01 &  2.62 &    5.43 &  0.12 &      25.80 &      26.87 &  1.04 \\
	G350.8162+0.5146 &   1.3 &  0.15 & 18.20 &  3.05 &    2.89 &  0.08 &      27.46 &     100.12 &  3.65 \\
	G351.1414+0.7764 &   1.4 &  0.12 & 18.82 &  4.10 &   10.42 &  0.24 &      51.15 &     201.26 &  3.93 \\
	G351.1510+0.7656 &   1.4 &  0.14 & 18.59 &  3.08 &    5.44 &  0.14 &      39.65 &     146.38 &  3.69 \\
	G351.1588+0.7490 &   1.4 &  0.12 & 21.30 &  3.64 &    7.84 &  0.19 &      42.00 &     364.07 &  8.67 \\
	G351.4981+0.6634 &   1.4 &  0.05 & 16.19 &  6.90 &  236.42 &  2.14 &      70.79 &     107.03 &  1.51 \\
	G351.5089+0.6415 &   1.3 &  0.10 & 16.40 &  4.16 &   12.83 &  0.26 &      41.91 &      71.50 &  1.71 \\
	G351.5290+0.6939 &   1.3 &  0.11 & 15.18 &  8.54 &   22.72 &  0.49 &      89.91 &      91.47 &  1.02 \\
	G351.5663+0.6068 &   1.3 &  0.15 & 13.69 &  5.37 &    7.51 &  0.22 &      73.17 &      41.42 &  0.57 \\
	G352.9722+0.9249 &   1.4 &  0.11 & 17.82 &  3.60 &    7.90 &  0.17 &      31.95 &     109.04 &  3.41 \\
	G353.0114+0.9828 &   1.3 &  0.14 & 18.93 &  2.24 &    3.58 &  0.10 &      31.05 &     135.10 &  4.35 \\
	G353.0195+0.9750 &   1.3 &  0.14 & 18.44 &  3.06 &    3.17 &  0.09 &      26.60 &     102.84 &  3.87    
	\enddata
	%\tablecomments{This table is available in its entirety in a machine-readable form in the online journal. A portion is shown here for guidance regarding its form and content.}
	%		\tablenotetext{}{\textit{Note ---} All the listed parameters, 
	%			except for $T_\mathrm{dust}$ and $N_\mathrm{H_2}$, are for the sample
	%			with distances.}
\end{deluxetable*}
	
	Our results differ substantially from those of
	\citet{2016ApJ...822...59S}.
	Our sample has higher median masses, surface densities,
	and volume densities. As a result there is much more overlap
	between the starless sample and the samples with ongoing star 
	formation in these properties (see Figures \ref{fig:density} and
	\ref{fig:histMass})
	than was found by \citet{2016ApJ...822...59S}. 
	There are small differences in the assumptions about opacity
	and the method of determining temperatures, as
	well as a slightly higher median distance, all of which tend
	to result in higher masses in our sample, but these are
	relatively small effects. The primary differences arise
	from the expanded longitude coverage, 
	the sample inclusion criteria, 
	the spatial resolution,
	and
	the source extraction method. 
	
	\citet{2016ApJ...822...59S} 
	considered sources with longitude exceeding 10 degrees, 
	while our range of $-60\deg < l < 60\deg$ includes the sources
	near the Galactic Center and some massive complexes in the
	fourth quadrant. If considering sources in $10\deg < l < 60\deg$ only, 
	the median mass
	of our HMSC candidates would decrease by 30\%, 
	but the median surface and column densities 
	don't change significantly.
	Svoboda et al. 
	included sources as weak as 100 mJy at 1.1 mm, while we 
	required the flux density at 870 \micron\ to be at least
	500 mJy beam$^{-1}$. Consequently, the median masses will be
	skewed higher in our sample (1019 \msun) than in the
	Svoboda 
	sample (228 \msun). 
	
	Perhaps the more puzzling difference between the samples
	is in the surface densities. Our sample has 234 
	sources with $\Sigma > 0.15$ g cm$^{-2}$, while
	Svoboda et al.
	have only 11 such sources. We examined the nature of these
	high surface density sources, focusing on the extreme cases
	with $\Sigma > 0.5$ g cm$^{-2}$. These fall into two distinctly
	different categories, as can be seen in Figure \ref{fig:Size_Sigma} where we plot
	size (angular and linear) versus $\Sigma$.
	About nine are large ($r_{\rm eq} > 1$ pc) clumps with very high
	$\Sigma > 0.5$ g cm$^{-2}$, and all lie in the Galactic Center
	region, which was excluded from the 
	Svoboda sample. 
	The others are compact, with angular sizes less than
	30\arcsec. These lie within larger clumps identified by 
	Svoboda et al.
	and are favored by the better spatial resolution of our sample,
	as well as the source finding method (Gaussclump) (section \ref{sec:data})
	which seeks small sources within extended structure 
	versus the method used by
	Svoboda 
	which is biased against splitting sources into smaller structures.
	These small $r_{\rm eq} < 0.15$ pc sources are candidates for dense 
	cores, rather than clumps.
	Some have masses exceeding 100 \msun. While follow-up studies
	will be needed, these are at least candidates for the 
	formation sites of individual, massive stars
	(see Section \ref{sec-cores}).
	
	\subsection{Possible High-mass Starless Cores}\label{sec-cores}

	As shown in Figure \ref{fig:Size_Sigma}, some of our sources
	have small physical sizes and high mass surface densities, 
	and probably represent cores embedded in
	clumps. Although there is no sharp definition,
	pc-scale entities are frequently referred to
	be clumps which would form a cluster of stars 
	and smaller structures with sizes of $0.01-0.3$ pc
	are commonly treated as cores which may form
	one or a group of stars 
	\citep[][and references therein]{2007ARA&A..45..339B,
		2015ApJ...804..141Z}. If we follow 
	\citet{2007ARA&A..45..339B} to adopt the 0.3 pc as 
	a threshold for discriminating between cores and clumps, about 4.3\% (20/463) of
	our HMSC candidates 
	with equivalent radii $<0.15$ pc would be high-mass starless 
	\textit{core} candidates. 
	
	The physical parameters 
	of these 20 possible starless cores are given Table \ref{tb-cores}.
	Fifteen of them have distances smaller than 2 kpc, and 14 reside
	in the well-known NGC 6334/6357 star-forming complex 
	\citep{2010A&A...515A..55R}.  Although further deep spectral imaging
	observations at a high-resolution are needed to check
	the quality of these cores, the large masses ($>26$) and 
	high densities ($\Sigma\ge0.08$ g cm$^{-2}$) still
	make them promising and have the potential
	to form massive stars. 

\section{Summary}\label{sec:conclusions}

	We have utilised data from multiple Galactic surveys to identify
	hundreds of high-mass starless clump candidates distributed 
	throughout the inner Galactic plane. The combination of 
	multiwavelength far-IR to submm continuum data enabled us to obtain some basic 
	parameters and we have compared these with those of
	clumps associated with methanol masers and \hii~regions. The 
	main findings in this work are summarized as follows:
	
	\begin{enumerate}
		\item From more than 10000 dense sources detected in the 
			ATLASGAL survey, a sample of 463 high-mass starless clump (HMSC) candidates
			were identified based on\textit{ Spitzer}/GLIMPSE, \textit{Spitzer}/MIPSGAL, 
			\textit{Herschel}/Hi-GAL and \textit{APEX}/ATLASGAL survey data.
			These clumps are not associated with any known star-forming indicators, the HMSC
			candidates represent a highly reliable catalog of 
			starless objects.
		\item Distances for all the 463 HMSC candidates were determined based on their systemic 
			velocities. Their distribution in
			Galactic longitude is similar to that observed in the whole sample of ATLASGAL 
			clumps, showing overdensities toward the Galactic center and several well
			known star-forming regions. While plotted on the face-on Milky Way, the majority 
			of the sources are located in spiral arms in the inner Galaxy, with
			Galactocentric distances less than 8.34 kpc.
		\item Some basic parameters were derived via fitting data at wavelengths from
			160 to 870 \micron~to modified blackbodies. These HMSC candidates
			have a median beam-averaged H$_2$ column density 
			of $4.4\times10^{22}$ cm$^{-2}$, a median 
			mass of 1019 \msun, a median luminosity of 1329 \lsun, and a median luminosity-to-mass
			ratio of 1.6 \lsun/\msun.
		\item More than 700 clumps associated with 6.7 GHz methanol masers and more than 400 clumps
			associated with \hii~regions were scrutinised using the same analysis techniques to enable us to carry out
			comparative diagnosis with the newly identified starless clumps. These comparison samples have median 
			size, mass and density similar to those of high-mass star-forming clumps.
			The HMSC candidates may be common entities but they are at an earlier evolutionary stage.
		\item All of the HMSC candidates except one fulfill the
			star-formation threshold proposed in \citet{2010ApJ...724..687L} 
			and \citet{2010ApJ...723.1019H}. In the mass-radius diagram, 
			more than 80\% (378/463) starless clumps are
			above the relationship for high-mass star formation 
			proposed by \citet{2010ApJ...723L...7K},
			About 97\% HMSC candidates have mass surface densities
			above the threshold defined in \citet{2014MNRAS.443.1555U},
			suggesting most of them have potential to form high-mass stars.
		\item Our HMSC candidates have median and mean dust temperatures of
			16.17 and 16.21 K, significantly colder than star-forming clumps and 
			consistent with other samples of starless clumps reported in the literature.
			The median luminosity-to-mass ratio of the HMSC candidates is as low as
			1.6 $L_\odot/M_\odot$. Further analysis shows that 
			these HMSCs are externally heated, suggesting that 
			these objects truly represent a very early phase of high-mass star formation. 
		\item Compared to the BGPS starless clumps in \citet{2016ApJ...822...59S}, 
			our HMSC candidates have larger masses and higher densities, which is 
			mainly due to the inclusion of the $-10\deg<l<10\deg$ region, the higher spatial resolution
			of ATLASGAL observations, and a different source extraction method.
		\item There are 20 HMSC candidates having equivalent radius 
		    $r_\mathrm{eq}<0.15$ pc. With small sizes,
			large masses, and high densities, they may represent a small sample of
			high-mass starless cores.
	\end{enumerate}
	
	We have identified the largest and most reliable high-mass starless 
	clump candidates ever reported. Distributed throughout the entire
	inner Galactic plane, these objects are ideal targets for further
	investigations on early stages of high-mass star formation. Follow-up
	observations toward these sources with 
	higher resolution and wider band coverage will contribute
	to hunting for high-mass starless cores, understanding 
	the fragmentation process, and revealing early chemistry.
	
\begin{acknowledgements}

    We are grateful to the anonymous referee for the 
    constructive comments that helped us improve this paper.
    This work is supported by the National Natural Science 
    Foundation of China through grants of 11503035, 11573036, 
    11373009, 11433008 and 11403040, the International 
    S\&T Cooperation Program of China through the grand of 
    2010DFA02710, the Beijing Natural Science Foundation 
    through the grant of 1144015, 
    the China Ministry of Science and Technology
    under State Key Development Program for Basic Research
    through the grant of 2012CB821800, and the Young Researcher 
    Grant of National Astronomical 
    Observatories, Chinese Academy of Sciences.
    KW is supported by grant 
    WA3628-1/1 through the DFG priority program 1573 
    ``Physics of the Interstellar Medium''.
    We thank Brain Svoboda for helpful discussions about 
    the differences between parameters in \citet{2016ApJ...822...59S}
    and this work.

    This research has made use of the NASA/IPAC Infrared Science Archive, 
    which is operated by the Jet Propulsion Laboratory, California Institute 
    of Technology, under contract with the National Aeronautics and Space 
    Administration. This research also has made use of the SIMBAD database,
    operated at CDS, Strasbourg, France.
    This work is based in part on observations made with the 
    \emph{Spitzer Space Telescope}, which is operated by the Jet 
    Propulsion Laboratory, California Institute of Technology under 
    a contract with NASA. This research made use of APLpy and Astropy 
    for visualization 
    and some analysis. APLpy is an open-source plotting package for 
    Python hosted at http://aplpy.github.com. And Astropy is a 
    community-developed core Python package 
    for Astronomy \citep{2013A&A...558A..33A}.

\end{acknowledgements}

\bibliography{HMSCs_Cat}

\end{CJK}
\end{document}